\documentclass[]{aa}
\usepackage{txfonts}
\usepackage{xcolor}
\usepackage{multirow}
\usepackage{subcaption} 
\usepackage{bigdelim}
\usepackage{longtable}
\usepackage{graphicx}
\usepackage{times}

\def\H2{H$_2$}

\setcitestyle{citesep={,}}
\begin{document}

\title{CH$_3$CN deuteration in the SVS13-A Class I hot corino, SOLIS XV}


\author{E. Bianchi \inst{1} 
\and
C. Ceccarelli \inst{1} 
 \and 
C. Codella \inst{2,1}
\and
A. L\'opez-Sepulcre \inst{3,1}
\and
S. Yamamoto \inst{4, 5}
\and
N. Balucani \inst{6,1,2}
\and 
P. Caselli \inst{7}
\and
L. Podio \inst{2}
\and
R. Neri \inst{3}
\and 
R. Bachiller \inst{8}
\and
C. Favre \inst{1} 
\and
F. Fontani \inst{2}
\and
B. Lefloch \inst{1}
\and
N. Sakai \inst{9}
\and
D. Segura-Cox \inst{7}
}

\institute{
Univ. Grenoble Alpes, CNRS, Institut de
Plan\'etologie et d'Astrophysique de Grenoble (IPAG), 38000 Grenoble, France
\and
INAF, Osservatorio Astrofisico di Arcetri, Largo E. Fermi 5,
50125 Firenze, Italy
\and
Institut de Radioastronomie Millim\'etrique, 300 rue de la Piscine, Domaine
Universitaire de Grenoble, 38406, Saint-Martin d'H\`eres, France
\and
Department of Physics, The University of Tokyo, 7-3-1, Hongo, Bunkyo-ku, Tokyo 113-
0033, Japan
\and
Research Center for the Early Universe, The University of Tokyo, 7-3-1, Hongo, Bunkyo-ku, Tokyo 113-0033, Japan
\and
Dipartimento di Chimica, Biologia e Biotecnologie, Via Elce di Sotto 8, 06123 Perugia, Italy
\and
Max-Planck-Institut f\"ur extraterrestrische Physik (MPE), 
Giessenbachstrasse 1, 85748 Garching, Germany
\and
Observatorio Astron\'omico Nacional (OAN-IGN), Alfonso XII 3,
28014, Madrid, Spain
\and
RIKEN Cluster for Pioneering Research, 2-1, Hirosawa, Wako-shi, 351-0198 Saitama, Japan
}

\offprints{E. Bianchi, \email{eleonora.bianchi@univ-grenoble-alpes.fr}}
\date{Received date; accepted date}

\authorrunning{Bianchi et al.}
\titlerunning{CH$_3$CN deuteration in the Class I object SVS13-A}
\abstract{Deuteration is a precious tool for investigating the origin and formation routes of interstellar complex organic molecules in the different stages of the star formation process.
 Methyl cyanide (CH$_3$CN) is one of the most abundant interstellar complex organic molecules (iCOMs);  it is of particular interest because it is among the very few iCOMs detected not only around protostars but also in protoplanetary disks. However, its formation pathways are not well known and only a few measurements of its deuterated isotopologue (CH$_2$DCN) have been made to date.}
{We studied the line emission from CH$_3$CN and its deuterated isotopologue CH$_2$DCN towards the prototypical Class I object SVS13-A, where the deuteration of a large number of species has already been reported. 
Our goal is to measure the CH$_3$CN deuteration in a Class I protostar, for the first time, in order to constrain the CH$_3$CN formation pathways and the chemical evolution from the early prestellar core and Class 0  to the evolved Class I stages.}
{We imaged CH$_2$DCN towards SVS13-A using the IRAM NOEMA interferometer at 3mm in the context of the Large Program SOLIS (with a spatial resolution of 1.$''$8 $\times$ 1.$''$2). 
The NOEMA images were   complemented by the CH$_3$CN and CH$_2$DCN spectra collected by the IRAM-30m Large Program ASAI, which provided an unbiased spectral survey at 3mm, 2mm, and 1.3mm. 
The observed line emission was analysed using local thermodynamic equilibrium (LTE) and non-LTE large velocity gradient (LVG) approaches.}
{The NOEMA/SOLIS images of CH$_2$DCN show that this species emits in an unresolved area centred towards the SVS13-A continuum emission peak, suggesting that methyl cyanide and its isotopologues are associated with the hot corino of SVS13-A, previously imaged via other iCOMs.
In addition, we detected 41 and 11 ASAI transitions of CH$_3$CN and CH$_2$DCN, respectively, which cover upper level energies (E$_{\rm up}$) from 13 to 442 K and from 18 K to 200 K.
The non-LTE LVG analysis of the CH$_3$CN lines points to a kinetic temperature of (140$\pm$20) K, a gas density n$_{\rm H_2}$ $\geq$ 10$^{7}$ cm$^{-3}$, and an emitting size of $\sim 0\farcs$3, in agreement with the hypothesis that CH$_3$CN lines are emitted in the SVS13-A hot corino.
The derived [CH$_2$DCN]/[CH$_3$CN] ratio is $\sim$9\%.
This value is consistent with those measured towards prestellar cores and a factor 2-3 higher than those measured in Class 0 protostars.}
{Contrarily to what expected for other molecular species, the CH$_3$CN deuteration does not show a decrease in SVS13-A with respect to measurements in younger prestellar cores and Class 0 protostars. 
Finally, we discuss why our new results suggest that CH$_3$CN was likely synthesised via gas-phase reactions and frozen  onto the dust grain mantles during the cold prestellar phase.}

\maketitle

\section{Introduction}\label{sec:intro}

In the context of the star formation process, Class I protostars (see e.g. \citealt{Andre1993, Caselli2012}, and references therein), with a typical age of 10$^{\rm 5}$ yr, are a bridge between the youngest Class 0 protostars (around 10$^{\rm 4}$ yr), where the bulk of the material feeding the protostar is still in the envelope, and the protoplanetary disks (around 10$^{\rm 6}$ yr). 
In addition, recently, ALMA showed gaps and rings in the distribution of millimetre dust grains in disks associated with less than 1 Myr, which are thought to be connected with the earliest phases of planet formation  \citep[e.g.][]{ALMA2015,Sheehan2017, Fedele2018, Segura-Cox2020}. 
These findings suggest that planet formation may occur already in the Class I stage. 
It is then promising to investigate the physical and chemical properties of the first stages of a Sun-like star and compare them with those found in our Solar System to reveal its early history.
In particular, it is not clear yet if the chemical complexity observed in our Solar System is, at least partially, inherited from the prestellar and protostellar phases or if instead there is a substantial chemical evolution. Measuring molecular deuteration (i.e. the abundance of the deuterated form of a molecule, [XD]) with respect to its undeuterated form ([XH]) at the different formation stages of a Sun-like star can help us to address this question.

Emission due to deuterated molecules is commonly observed in all the evolutionary stages, from the prestellar core phase until the formation of a Sun-like star (e.g. \citealt{Caselli2012}, and references therein). 
 These observations can be used to efficiently trace the chemical evolution along the star formation process, as suggested by water deuteration, which decreases with time from protostars to the bodies of our Solar System \citep{Ceccarelli2014,Furuya2017,Jensen2021}.
More specifically, deuteration is an important  tool for the study of hot corinos, which are compact regions around   protostars (< 100 au) where the temperature is high enough ($>$ 100 K) to sublimate the molecules frozen onto dust mantles in the gas phase. 
Given the high temperature in the hot corino, the deuteration there is a fossil, a precious record of the processes that occurred at the time of the dust mantle formation when the source was in cold conditions (e.g. \citealt{Taquet2012b, Aikawa2012, Codella2012, Bianchi2019b}, and references therein). 
Of particular interest is deuteration of interstellar complex organic molecules \citep[iCOMs;][]{Ceccarelli2017,Herbst2009}, which are the building blocks contributing to prebiotic chemistry. Since the deuteration process is very sensitive to the gas physical conditions, measurements of iCOM deuteration provide important constraints on their origin and formation pathways \citep[e.g.][]{Coutens2016,Skouteris2017,Taquet2019,Manigand2019,Agundez2021}.
Regarding Class I objects, few sources have been observed using D-species \citep[e.g.][]{LeGal2020}. 
Among them, only the SVS13-A hot corino was extensively investigated using several molecular tracers by \citet{Codella2016b} and \citet{Bianchi2017a, Bianchi2019b}.
These authors showed that H$_2$CO, H$_2$CS, and HC$_3$N have a deuteration similar to that measured towards Class 0 protostars, 
while  CH$_3$OH presents a molecular deuteration that seems to decrease by at least one order of magnitude. 
We definitely need to measure the molecular deuteration in other species to obtain a more complete and hopefully coherent picture, and to be able to  efficiently use astrochemical models (see e.g. \citealt{Aikawa2012, Taquet2019}).
A step ahead in the comprehension of how deuteration evolves during the star formation process can be obtained using CH$_{\rm 3}$CN.
This species can be considered one of the most abundant iCOMs in low-mass star-forming regions.
It is also one of the few iCOMs detected in Class 0/I {and} protoplanetary disks  \citep{Codella2009, Oberg2014, Oberg2015, Bergner2018, Loomis2018, Taquet2015, Belloche2020, Yang2021}.
In addition, CH$_{\rm 3}$CN has been detected in comets, including towards 67/P in the context of the Rosetta mission \citep{LeRoy2015, Altwegg2019}.
On the other hand, measurement of both CH$_{\rm 3}$CN and CH$_{\rm 2}$DCN in young solar  analogues have been reported so far only towards a limited number of objects \citep{Calcutt2018, Taquet2019, Agundez2019, Cabezas2021, Yang2021, Nazari2021}.
However, to our knowledge, no specific study on the CH$_{\rm 3}$CN deuteration has been performed yet.

\textbf{The SVS13-A Class I laboratory:} SVS13-A is a young star located in the well-known NGC1333 cluster in the Perseus region at a distance of 299 $\pm$ 14 pc, as recently measured by the mission {\it Gaia}\footnote{\url{http://www.esa.int/Science_Exploration/Space_Science/Gaia_overview}}\citep{Zucker2018}.
The source has been subject of a large number of observational campaigns in
different spectral windows (see e.g. \citealt{Chini1997, Bachiller1998, Looney2000, Chen2009, Tobin2016, Lefloch2018, Ceccarelli2017, Maury2019, Diaz2021}, and references therein).
SVS13-A has a bolometric luminosity $\sim$ 50 $L_{\rm \odot}$ and
a bolometric temperature $\sim$ 188 K, is classified as a Class I source (at least 10$^5$ yr, e.g. \citealt{Chini1997}), and   is in turn a close binary source (VLA4A, VLA4B with 0$\farcs$3 separation; \citealt{Anglada2000}).
The SVS13-A system is still associated with a large envelope \citep{Lefloch1998a}, and it is driving an extended molecular outflow \citep{Lefloch1998a, Codella1999}, as well as the Herbig-Haro chain 7--11 \citep{Reipurth1993}.
More recently, a chemically rich hot corino has been detected towards SVS13-A using deuterated water and iCOM line emission \citep{Codella2016b, Desimone2017, Bianchi2019a,Belloche2020, Yang2021}.
The hot corino was imaged by \citet{Desimone2017} using HCOCH$_2$OH (glycolaldehyde) emission lines and its size was   estimated to be about 90 au (300 mas).
In addition, \citet{Lefevre2017} suggests that the chemical richness observed towards SVS13-A is associated with the VLA4A object. This has been confirmed by \citet{Diaz2021} using high-angular resolution observations.
Very recently, several studies have been focused on the molecular deuteration of SVS13-A, using HDO, CH$_2$DOH, HDCO, D$_2$CO, HDCS, and DC$_3$N \citep{Codella2016b,Bianchi2017a, Bianchi2019b}.
These studies show some conflicting results:  they do not suggest a dramatic decrease in deuteration in the observed molecules with respect to the earlier stages represented by the Class 0 protostars, with the exception of methanol.
However, no firm conclusion could be drawn, calling for a more extensive study of molecular deuteration in other species.

We present here the first study of CH$_3$CN deuteration in a Class I protostar. The paper is  organised as follows. In Sect. 2 we describe the observations. In Sect. 3 we present our results on the CH$_2$DCN spatial distribution and we derive the gas properties (excitation temperature, column density) for CH$_3$CN and CH$_2$DCN, using a non-local thermodynamic equilibrium (LTE) large velocity gradient (LVG) analysis and a LTE rotational diagram analysis, respectively. We discuss in Section 4 the obtained CH$_3$CN deuteration, and we compare it with measurements in other sources. We discuss the possible chemical formation routes in light of our results.  Finally, we present our conclusions in Sect. 5.

\section{Observations} \label{sec:observations}
In this paper we analyse the observations from two complementary datasets. The observations were taken towards SVS13-A, at the coordinates $\alpha_{\rm J2000}$ = 03$^{\rm h}$ 29$^{\rm m}$ 03$\fs$76, $\delta_{\rm J2000}$ = +31$\degr$ 16$\arcmin$ 03$\farcs$0.
The first dataset was obtained with the IRAM/NOEMA interferometer\footnote{\url{http://www.iram-institute.org/}} as part of the Large Program Seeds of Life in Space\footnote{\url{http://solis.osug.fr/}} \citep[SOLIS;][]{Ceccarelli2017} and provides high spatial resolution maps of two lines from singly deuterated methyl cyanide (CH$_2$DCN),  $5_{\rm 1,4}$--$4_{\rm 1,3}$ and  $6_{\rm 1,6}$--$5_{\rm 1,5}$, whose spectroscopic parameters are reported in Table 1.
The second dataset was obtained with the IRAM-30m$^2$ single-dish telescope as part of the Large Program Astrochemical Survey At Iram\footnote{\url{http://www.oan.es/asai/}} \citep[ASAI;][]{Lefloch2018} and contains several lines from methyl cyanide and its singly deuterated isotopologue.

\subsection{NOEMA/SOLIS} \label{subsec:obs-solis}

The observations were obtained during two tracks of 1.9 hr and 6.4 hr using nine antennas in A configuration on March 16 and March 24, 2018.
The shortest and longest projected baselines are 64 and 760 m, respectively. 
The field of view is about 60$\arcsec$, while the largest angular scale (LAS) is about 4$\arcsec$. 
We used the Polyfix correlator, which covered two frequency ranges,  about 80--88 and 96--104 GHz, respectively, with a spectral resolution of 2.0 MHz ($\sim$6--7 km/s). 
The calibration was performed following the standard procedures, using GILDAS-CLIC\footnote{\url{http://www.iram.fr/IRAMFR/GILDAS}}. 
The bandpass was calibrated on 3C84, while the absolute flux was calibrated by observing LkH$\alpha$101, MWC249, and the phase using 0333+321.
The final uncertainty on the absolute flux scale is $\leqslant$10$\%$. 
The phase rms was $\leqslant$50$^\circ$, the typical precipitable water vapour (pwv) about 5-15 mm, and the system temperature about 50-150 K. The data were self-calibrated in phase only, and the solutions applied to the data spectral cube.
 Line images were produced by subtracting the continuum image (derived using line-free channels), using natural weighting, and restored with a clean beam of 1.$''$8 $\times$ 1.$''$2 (PA= 39$^\circ$). 
The rms noise in the broad-band cubes at the CH$_2$DCN frequencies is 0.7 mJy/beam.

\subsection{IRAM/ASAI} \label{subsec:obs-asai}

The reported observations were obtained during several runs between 2012 and 2014, as described by \citet{Lefloch2018}.
They provide an unbiased spectral survey towards SVS13-A of the 3 mm (80--116 GHz), 2 mm (129--173 GHz), and 1.3 mm (200--276 GHz) bands accessible with IRAM-30m. 
In this work, we report and analyse all the CH$_3$CN, and CH$_2$DCN lines falling in these bands. 
The telescope half power beam width (HPBW) ranges from $\simeq$ 9$\arcsec$ at 276 GHz to $\simeq$ 30$\arcsec$ at 80 GHz. 
The observations were acquired in wobbler switching mode with a 180$\arcsec$ throw.
The  broad-band EMIR receivers were used, connected to the FTS200 backends, which provide a spectral resolution of 200 kHz, corresponding to channels of 0.7 (at 3 mm) to 0.2 km/s (1 mm).
The pointing error was found to be less than 3$\arcsec$, while the uncertainty on the calibration is from $\sim$ 10\% (3 mm) to $\sim$ 20\% (1 mm).
At the frequencies of the observed CH$_3$CN and CH$_2$DCN lines, the rms noise (in $T_{\rm MB}$ scale) ranges from 2 (3 mm) to 35 mK (1 mm). 


\section{Results} \label{sec:results}

\subsection{NOEMA/SOLIS results: CH$_2$DCN emission maps.} \label{subsec:results-solis}

Figure \ref{fig:solis-maps} shows the spatial distribution of the dust continuum emission at 3 mm. 
In addition to SVS13-A, two other Class 0 objects,  VLA3 and SVS13-B, are detected in the primary beam of NOEMA observations. 
All measured positions are in agreement with those previously derived at millimetre wavelengths \citep[e.g.][]{Maury2019}:  
SVS13-A: 03$^h$29$^m$03$^s$.757, +31$^\circ$16$'$03$\farcs$74; VLA3: 03$^h$29$^m$03$^s$.386, +31$^\circ$16$'$01$\farcs$56;
SVS13-B: 03$^h$29$^m$03$^s$.064, +31$^\circ$15$'$51$\farcs$50.
The continuum towards SVS13-A has a roundish shape with a diameter of about 4$''$, corresponding to about 1200 au.
This emission very likely probes the dense and warm inner envelope surrounding the central protostar.
In the same figure we show the spatial distribution of the emission from the two CH$_2$DCN lines, whose spectroscopic parameters are reported in Table \ref{table:CH3CN-lines}.
Figure \ref{fig:SOLIS-CH3CN-spectra} shows the spectra extracted at the peak position.
Contrarily to the continuum, the line emission is only detected towards SVS13-A and it is unresolved by the NOEMA beam at 3mm;  it has a  diameter of less than about 1$\farcs$5 or 450 au.
This suggests that the CH$_2$DCN lines trace the inner part of the envelope and/or the hot corino of SVS13-A, where icy dust grain mantles sublimate, releasing their content in the gas phase. The hypothesis that the emitting region of methyl cyanide and its isotopologue is the hot corino region is further confirmed by the CH$_3$CN non-LTE analysis (Sect. \ref{subsec:nonLTE}).

\subsection{IRAM/ASAI results: CH$_{\rm 3}$CN and CH$_{\rm 2}$DCN} \label{subsec:results-asai}

\begin{figure*}[h!]
\begin{center}
\includegraphics[angle=90,width=18cm]{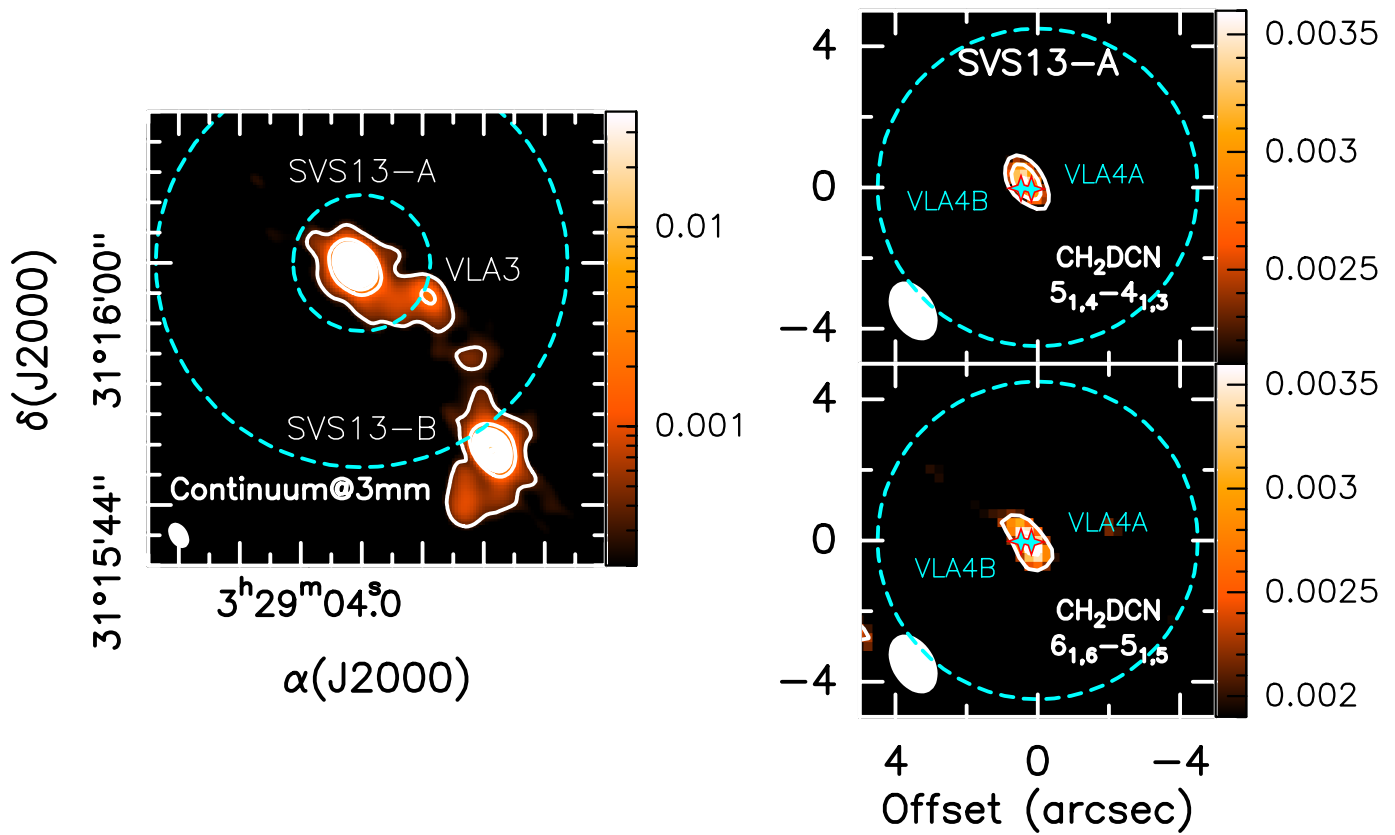}
\caption{NOEMA/SOLIS images of the continuum and CH$_2$DCN line emission towards SVS13-A.
\textit{Left panel}: Dust continuum emission at about 90 GHz (see Sect. \ref{sec:observations}). The intensity scale is in Jy beam$^{-1}$.
The first contour and the steps are 5$\sigma$ (1$\sigma$ = 7 $\mu$Jy/beam) and 20$\sigma$, respectively.
The filled ellipse in the bottom left corner shows the synthesised beam (HPBW): 1.$''$8 $\times$ 1.$''$2 (PA= 39$^\circ$).
The three objects in the primary beam, SVS13-A, SVS13-B, and VLA3, are labelled. 
The cyan dashed circles indicate the minimum (9$\arcsec$) and maximum (27$\arcsec$) IRAM-30m HPBW in the 1 and 3 mm bands, respectively (see Sect. \ref{sec:results}).
\textit{Right panels}: Zoomed-in images of the central region where the two CH$_2$DCN lines emit. The intensity  scale is Jy beam$^{-1}$ km s$^{-1}$.
The first contours and steps are 3$\sigma$ (2.1 mJy beam$^{-1}$ km s$^{-1}$) and 1$\sigma$, respectively.
The synthesised beam is the same as for  the continuum map. 
The two components of the SVS13-A binary system, VLA4A and VLA4B,  are 0$\farcs$3 apart \citep{Anglada2000} and are labelled.
The cyan dashed circle shows the 9$\arcsec$ IRAM-30m HPBW.}
\label{fig:solis-maps}
\end{center}
\end{figure*}

\begin{figure}
\begin{center}
\includegraphics[angle=0,width=8cm]{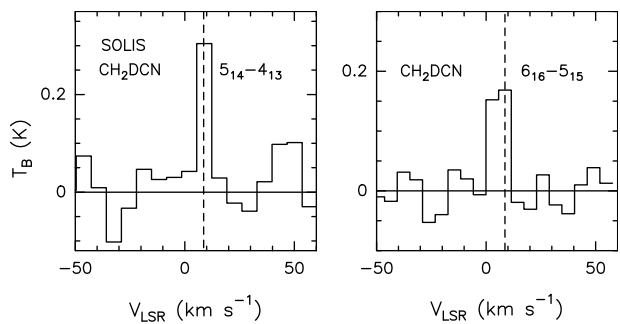} 
\caption{Spectra of the CH$_{\rm 2}$DCN lines observed with NOEMA/SOLIS and whose spatial distribution is shown in Fig. \ref{fig:solis-maps}. 
The spectra are extracted at the line peak position and reported in brightness temperature ( T$_{\rm B}$) scale. 
The vertical dashed lines give the ambient LSR velocity \citep[+8.6 km/s:][]{Chen2009}.}
\label{fig:SOLIS-CH3CN-spectra}
\end{center}
\end{figure}

The full coverage of the 3, 2, and 1 mm bands with the IRAM-30m  antenna enabled the detection of 41 lines from CH$_3$CN and 7 lines from CH$_2$DCN covering 11 transitions. 
Line identification was   performed using the Jet Propulsion Laboratory (JPL\footnote{https://spec.jpl.nasa.gov/}; \citealt{Pickett1998}) and Cologne Database for Molecular Spectroscopy (CDMS\footnote{https://cdms.astro.uni-koeln.de/}; \citealt{Muller2005}) molecular data bases, and double-checked with the GILDAS Weeds package \citep{Maret2011}. For CH$_{\rm 3}$CN, all the detected lines have a signal-to-noise ratio higher than 5$\sigma$, while for CH$_{\rm 2}$DCN four of the detected lines have a signal-to-noise ratio higher than 5$\sigma,$ while other three have a signal-to-noise ratio between 5$\sigma$ and 3$\sigma$.
Since the beam is a function of the frequency, lines from different bands probe different regions, as shown in Fig. \ref{fig:solis-maps}.
While the beams in the 3 and 2 mm band also intercept  emission from VLA3 and SVS13-B, the emission from lines lying in the 1 mm band is dominated by SVS13-A.
Finally, we carefully checked that these lines are not blended with emission due to other species.

\begin{figure*}
\begin{center}
\includegraphics[angle=0,width=16cm]{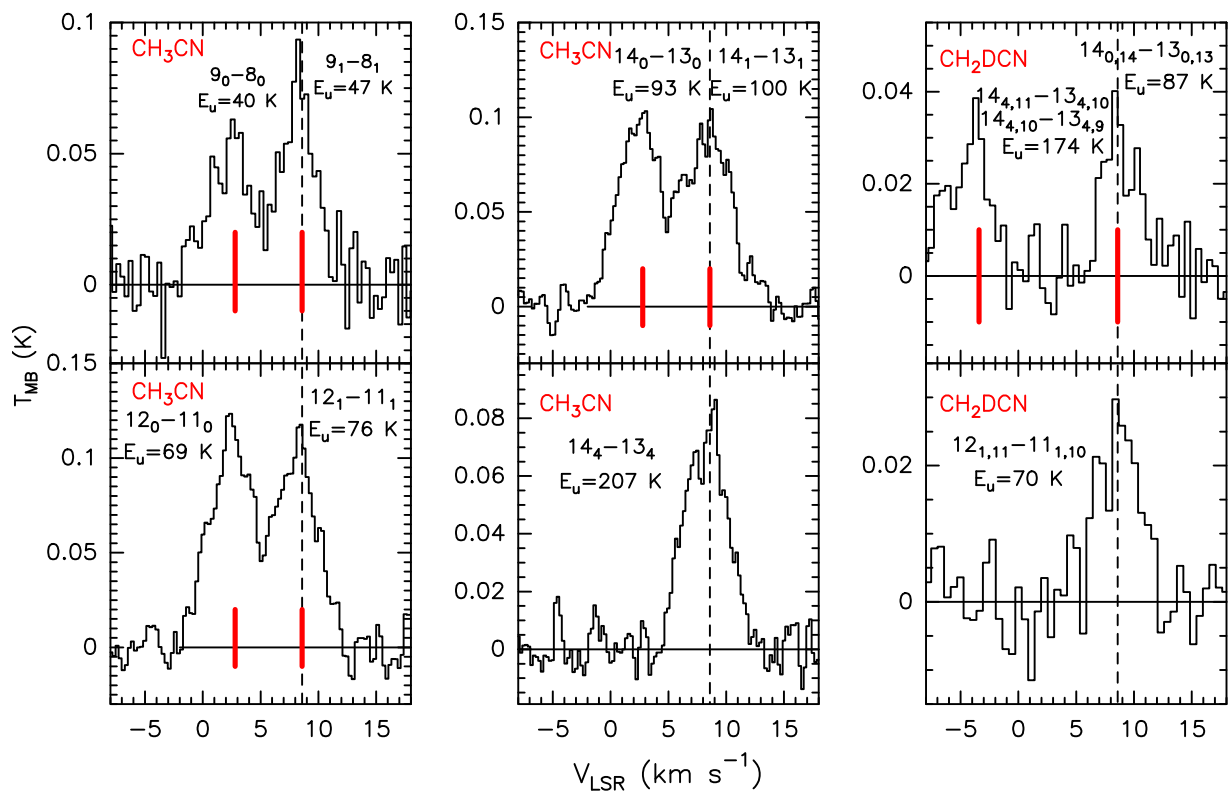} 
\caption{Examples of CH$_{\rm 3}$CN and CH$_{\rm 2}$DCN lines observed by ASAI, in T$_{\rm MB}$ scale (not corrected for beam dilution).
The spectroscopic data are listed in Table \ref{table:CH3CN-lines}. 
The vertical dashed line gives the ambient LSR velocity (+8.6 km s$^{-1}$, \citealt{Chen2009}). 
If multiple lines are present in the spectral window, the vertical red lines indicate their positions.}
\label{Fig:spectra-CH3CN}
\end{center}
\end{figure*}

All the detected lines were fitted using the GILDAS$^5$ CLASS package and assuming Gaussian profiles.
All the lines peak at velocities close to the SVS13-A systemic velocity v$_{\rm sys}$ = +8.6 km s$^{-1}$ \citep[e.g.][]{Chen2009}. 
Figure \ref{Fig:spectra-CH3CN} shows a representative sample of the detected lines, while Table \ref{table:CH3CN-lines} reports the list of all detections with their spectroscopic and derived line parameters, namely the integrated line intensity ($I_{\rm int}$), the line full width at half maximum (FWHM),  the line peak velocity ($V_{\rm peak}$), and the main beam temperature (in $T_{\rm MB}$ scale). The whole ASAI CH$_3$CN and CH$_2$DCN spectra are shown in the Appendix (see Fig. \ref{Fig:spectra+model1} and Fig. \ref{Fig:spectra+model2}). 

\begin{figure}
\begin{center}
\includegraphics[angle=0,width=9cm]{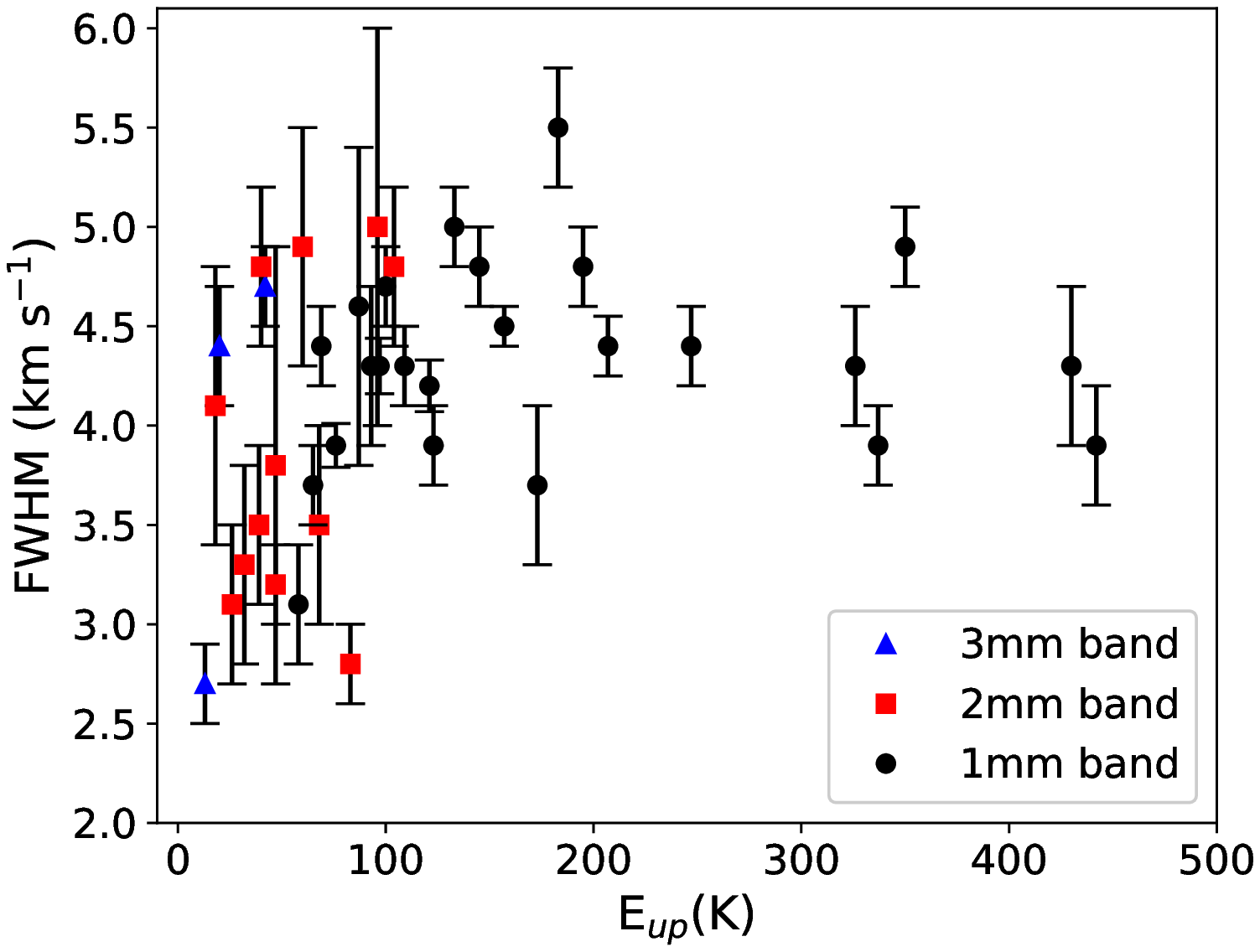} 
\caption{Line widths (FWHMs) of the CH$_{\rm 3}$CN lines detected with the ASAI observations (Table \ref{table:CH3CN-lines}) as a function of the upper level energy of the transitions ($E_{\rm up}$). The different colours are for the lines observed in different IRAM-30m bands:
blue (3mm), red (2mm), and black (1.3mm).}
\label{Fig:FWHM}
\end{center}
\end{figure}

The detected CH$_3$CN lines cover the 5$_{\rm K}$--4$_{\rm K}$ to 14$_{\rm K}$--13$_{\rm K}$ spectral ladders and their upper level energies (E$_{\rm up}$)  range from 13 K to 442 K.
They are detected in all  three of the ASAI bands, even though most of them (24/41) are detected at 1 mm.
The observed CH$_3$CN line emission could be associated with the relatively extended cold envelope and the sum of SVS13-A, SVS13-B, and VLA3 in the bands at 3 and 2 mm, whereas the band at 1 mm encompasses only SVS13-A.
A different origin of the line emission in the three bands is also suggested by the FWHM  of the CH$_3$CN lines which ranges from 2.7 km s$^{-1}$ 
to 5.5 km s$^{-1}$.
In order to better constrain the spatial origin of the CH$_3$CN line emission in the ASAI dataset, we plotted the line FWHM as a function of the line upper level energy in Fig. \ref{Fig:FWHM}.
A trend is evident: while the highest excitation lines (up
to more than 400 K)  always show
line widths  larger than $\sim$ 4 km s$^{-1}$, 
the lowest excitation lines show a large spread in the line width distribution, with 
values down to 2.7 km s$^{-1}$. This suggests that low-excitation lines can be contaminated by the extended cold envelope \citep[see e.g.][]{Ceccarelli2003}.
We note that all the lines observed in the 1 mm band have upper level energy higher than 60 K and larger FWHMs, indicating that they are dominated by emission from the SVS13-A hot corino.
The CH$_2$DCN lines are only detected in the 1 mm band and they all have high upper level energies, between 70 and 200 K.
Therefore, they are very likely exclusively emitted in the hot corino of SVS13-A (see also \S ~\ref{subsec:results-solis}).
The two CH$_2$DCN lines detected with NOEMA/SOLIS are not detected by the ASAI survey. We verified that this is due to beam dilution (the comparison between SOLIS and ASAI spectra is shown in Fig. \ref{Fig:asai-solis}).


\subsection{CH$_3$CN non-LTE analysis} \label{subsec:nonLTE}

In order to estimate the physical conditions of the methyl cyanide emitting gas, namely gas temperature, density, and CH$_3$CN column density, we used the non-LTE LVG code \rm{grelvg} described in \cite{Ceccarelli2003}. 
We used the collisional coefficients of CH$_3$CN with H$_2$, scaled from He, computed by \citet{Green1986} between 20 and 140 K and for J$\leq$25. 
The coefficients were retrieved from the LAMDA\footnote{\url{https://home.strw.leidenuniv.nl/~moldata/}} database \citep{Schoier2005}, where values are extrapolated for temperatures higher than 140 K.
We assumed a semi-infinite slab geometry to compute the line escape probability \citep{Scoville1974} and adopted a line width equal to 4.5 km s$^{-1}$, as indicated by the observations
(see Table \ref{table:CH3CN-lines} and Fig. \ref{Fig:FWHM}).

We consider for our analysis the CH$_3$CN lines in the 1 mm band, specifically with frequencies higher than 202 GHz, in order to avoid possible emission from SVS13-B, which falls in the 27$\arcsec$ beam at 3mm, and to probe only the gas in the hot corino of SVS13-A.
We ran a large grid of models ($\sim$3000) varying the kinetic temperature T$_{\rm kin}$ from 50 to 200 K, the H$_2$ density n$_{\rm H_{\rm 2}}$ from 10$^5$ to 10$^8$ cm$^{-3}$, and the CH$_3$CN column density N(CH$_3$CN) from 10$^{15}$ to 10$^{18}$ cm$^{-2}$.
We then fitted the measured CH$_3$CN velocity-integrated line intensities via comparison with those predicted by the \rm{grelvg} model, leaving T$_{\rm kin}$, n$_{\rm H_{\rm 2}}$, N(CH$_3$CN), and the emitting size as free parameters. 
We note that, in the fitting, we added 20\% of calibration uncertainty to the statistical errors listed in  Table \ref{table:CH3CN-lines}.

\begin{figure}[h]
\begin{center}
\includegraphics[angle=0,width=8.4cm]{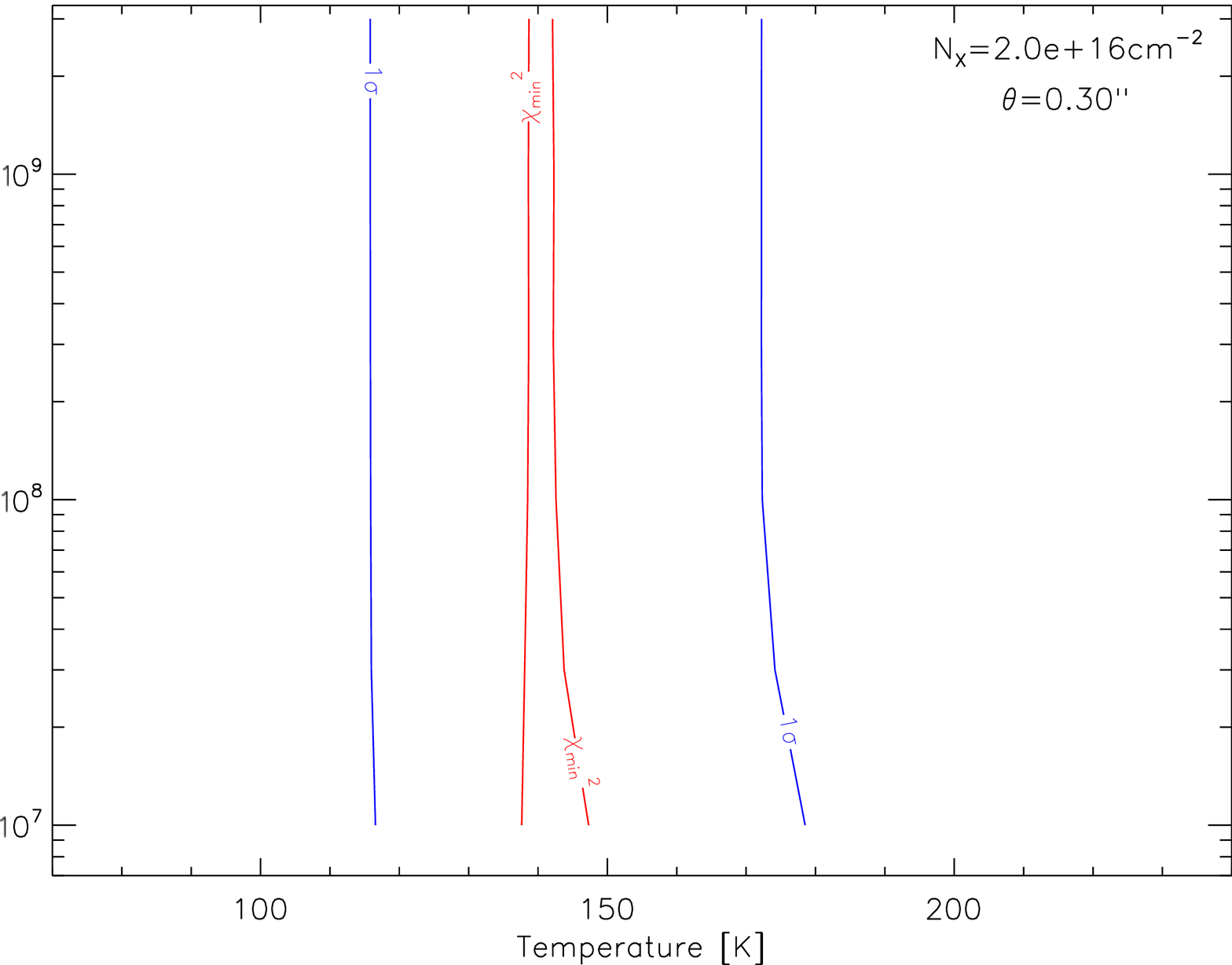} 
\includegraphics[angle=0,width=8.4cm]{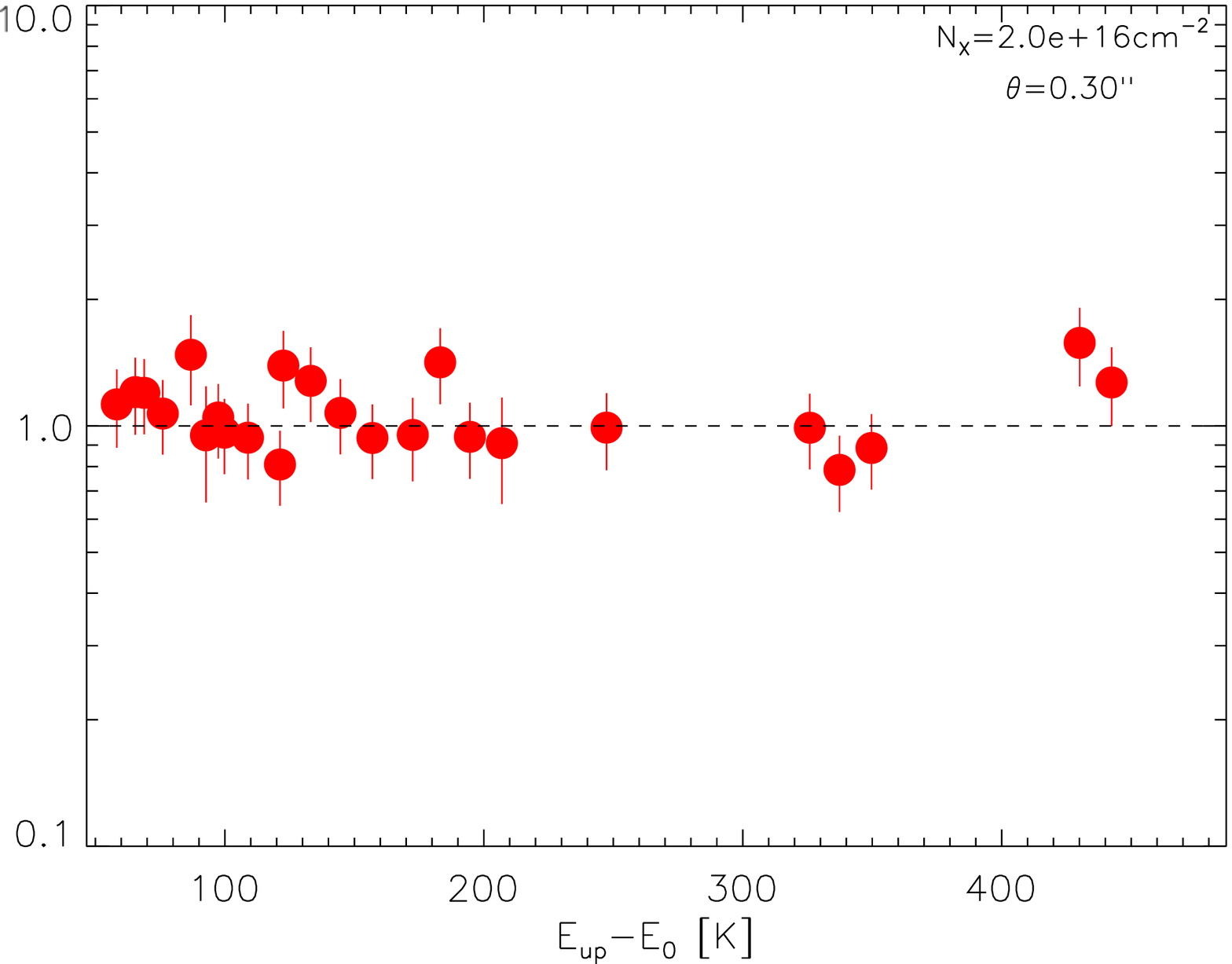} 
  \caption{LVG modelling of CH$_3$CN. {\it Upper panel:} $\chi^2$ contour plot as a function of  density (n$_{\rm H_{\rm 2}}$ on y-axis) and temperature (x-axis) at the best fit position with respect to the CH$_3$CN column density (N$_{\rm CH_3CN}$ = 2 $\times$ 10$^{16}$ cm$^{-2}$) and an emitting size (0$\farcs$3 in diameter). {\it Lower panel:} Observed to theoretical intensities as a function of the upper level energy of the line, at the best fit position, i.e. with a CH$_3$CN column density N(CH$_3$CN) = 2 $\times$ 10$^{16}$ cm$^{-2}$, an emitting size 0$\farcs$3 in diameter, kinetic temperature T$_{\rm kin}$=140 K, and volume density  n$_{\rm H_{\rm 2}}$ higher than 10$^7$ cm$^{-3}$.}
  \label{Fig:bestfit}
  \end{center}
\end{figure}

The best fit ($\chi^2_{\rm{ reduced}}$ = 0.79) of the data is obtained with N(CH$_3$CN) = $2 \times 10^{16}$ cm$^{-2}$ and an emitting size of 0$\farcs$3 in diameter.
The reduced $\chi^2_{\rm{ reduced}}$ is less than unity for N(CH$_3$CN) between 5 $\times$ 10$^{15}$ and 5 $\times$ 10$^{16}$ cm$^{-2}$.
Figure \ref{Fig:bestfit} (upper panel) shows the density--temperature contour plot of the $\chi^2$ at the best fit of N(CH$_3$CN) and size.
The kinetic temperature is very well constrained at  (140 $\pm$ 20) K, and we can determine the density, $\geq 10^7$ cm$^{-3}$, being the levels LTE populated.
The vast majority of the lines are optically thick, with optical depths of up to 5.
Only three lines, namely the 12$_{\rm 6}$--11$_{\rm 6}$, 12$_{\rm 5}$--11$_{\rm 5}$, and 14$_{\rm 7}$--13$_{\rm 7}$ transitions, have optical depths of less than unity (the lowest value being 0.3).
Figure \ref{Fig:bestfit} (lower panel) also shows the ratio of the observed to predicted intensity as a function of the upper level energy of the line. All the observed velocity-integrated line intensities are very well reproduced by the LVG  modelling, with no line more than 2$\sigma$ away from the best fit. 

The high density obtained from the LVG modelling ensures that LTE is a good approximation for CH$_3$CN. For this reason, we generated LTE synthetic spectra, using the gas temperature and column density and the associated errors, as derived from the best LVG model. The synthetic spectra, generated using the GILDAS Weeds package, are overlaid on the observations in Fig. \ref{Fig:spectra+model1}. The comparison shows a reasonable agreement considering the observed FWHMs distribution (see Fig. \ref{Fig:FWHM} and Table \ref{table:CH3CN-lines}). 
We note that some  low-K transitions show a hint of emission from a narrow component (see Fig. \ref{Fig:spectra+model1}), likely indicating the presence of emission from a colder extended envelope. 
Further observations mapping the large-scale envelope in both CH$_{\rm 3}$CN and its deuterated form are required to correctly disentangle the different contributions.

\subsection{CH$_2$DCN column density} \label{subsec:CH2DCN-LTE}

Since no collision coefficients exist for the CH$_2$DCN--H$_2$ system, in order to derive the CH$_2$DCN column density we carried out a rotation diagram LTE analysis. We used only four lines detected in ASAI, which consist of a single transition, namely: 12$_{\rm 1,11}$--11$_{\rm 1,10}$, 13$_{\rm 2,11}$--12$_{\rm 2,10}$, 14$_{\rm 0,14}$--13$_{\rm 0,13}$, and 14$_{\rm 1,13}$--13$_{\rm 1,12}$.
We assumed that CH$_2$DCN is emitted from the same region that emits in CH$_3$CN, namely the hot corino region, as suggested by the SOLIS maps (see Sect. \ref{subsec:results-solis}).
Consequently, we used an emitting size of 0$\farcs$3 in diameter and a temperature of 140 K. The derived column density is $2.4^{+0.3}_{-0.4} \times 10^{15}$ cm$^{-2}$. The error considers a range of assumed temperatures between 120 and 160 K. 
Varying the assumed temperature from 50 to 200 K, the CH$_2$DCN column density is always higher than 1.5 $\times$ 10$^{15}$ cm$^{-2}$. 
Under these conditions all the lines are predicted to be optically thin, with opacities lower than 0.2. The results do not change if we include in the rotation diagram the two SOLIS lines.
The results of the analysis are summarised in Table \ref{tab:RD}. LTE synthetic spectra are generated for the best fit model and the associated error, and they are overlaid on the observed spectra in Fig. \ref{Fig:spectra+model2} for CH$_2$DCN.  

\begin{table}
\caption{Results of the CH$_3$CN  and CH$_2$DCN analysis (see text). 
The columns list  (1) the species; (2) the kinetic temperature, $T_{\rm kin}$, derived from the CH$_3$CN non-LTE analysis; (3) the column densities, $N_{\rm xt}$, derived assuming the same temperature for both species; (3) the deuteration.}
\begin{tabular}{lccc}
\hline
Species & T$_{\rm kin}$ & N$_{\rm x}$$^a$  & [XD]/[XH] \\
         &  (K)   & (cm$^{-2}$)  & (\%) \\
\hline
CH$_3$CN  & 140 (20) & (5  -- 50) ~$\times 10^{15}$ & -- \\
CH$_2$DCN & 120-160$^b$      & (2.0 -- 2.7) ~$\times 10^{15}$ & 4 -- 54\\

\hline 
\end{tabular}\\
$^a$ Derived assuming an emitting size of 0$\farcs$3.\\
$^b$ Assumed from CH$_3$CN.
\label{tab:RD}
\end{table}


\subsection{CH$_3$CN deuteration in SVS13-A} \label{subsec:disc-deut-SVS13a}

A first estimate of CH$_3$CN deuteration was   obtained by computing the ratio of the CH$_2$DCN to the CH$_3$CN column densities, assuming the same temperature for both species. 
With this method we derived a range of CH$_3$CN deuteration between 4$\%$ and 54$\%$ (see Table \ref{tab:RD}), with a best value of 10$\%$.

We also derived the CH$_3$CN deuteration using a second method,  dividing the intensities of lines with the same quantum number $J$ in the two species and similar upper-level energies, following the same procedure adopted for HC$_3$N and H$_2$CS in \citet{Bianchi2019b} (see also \citealt{Kahane2018}).
In general, this method allows a straightforward derivation of the abundance ratio, which does not depend on the assumed temperature, and therefore is affected by lower uncertainty. 

While CH$_3$CN is a symmetric top molecule, CH$_2$DCN is a near prolate asymmetric top molecule.
This means that the CH$_3$CN transitions are described by two rotational quantum numbers: the total angular momentum, $J$, and its projection on the symmetry axes, $K$. For CH$_2$DCN there is no symmetry axis and the rotational quantum numbers are denoted  $J_{K_{-1}K_{1}}$;  in particular, for prolate rotors the quantum state is $J_{K_{-1}}$. 
Therefore, for each CH$_3$CN transition, we have two corresponding CH$_2$DCN transitions, called the K-type doublet, with the same $J$ and $K_{-1}$, if $K_{-1}$ $\neq$ 0, and only one CH$_2$DCN transition if $K_{-1}$ = 0.
For example, the intensity of the CH$_2$DCN line at 243.0512 GHz, composed of the two transitions 14$_{\rm 4,11}$--13$_{\rm 4,10}$ and 14$_{\rm 4,10}$--13$_{\rm 4,9}$ (E$_{\rm up}$ = 174 K), is divided for the intensity of the 14$_{\rm 4}$--13$_{\rm 4}$ CH$_3$CN line at 257.4481 GHz (E$_{\rm up}$= 207 K).
With the same method we calculated the CH$_2$DCN/CH$_3$CN ratio using the 14$_{\rm 0,14}$--13$_{\rm 0,13}$ CH$_2$DCN transition at 243.0415 GHz (E$_{\rm up}$= 87 K).
For the other three CH$_2$DCN lines, the 12$_{\rm 1,11}$--11$_{\rm 1,10}$ transition (E$_{\rm up}$= 70 K), the 13$_{\rm 2,11}$--12$_{\rm 2,10}$ transition (E$_{\rm up}$= 97 K), and the 14$_{\rm 1,13}$--13$_{\rm 1,12}$ transition (E$_{\rm up}$= 93 K), only one of the K-doublet transitions is exploitable for the analysis since the other one is blended (see Table \ref{table:CH3CN-lines} and Fig. \ref{Fig:spectra+model2}). 
In this case, we multiplied the line intensity by two since we expected the same intensity from the K-doublet lines.

We note that, given the presence of a pair of identical hydrogen nuclei, CH$_2$DCN presents two sets of nuclear-spin functions corresponding to {\it ortho} and {\it para} states: three functions for {\it ortho} and one for {\it para}. However, these {\it ortho} and {\it para} nuclear-spin functions do not couple to the specific rotational wave functions. Since the rotation motion cannot interchange the two hydrogen nuclei for CH$_2$DCN, the restriction of the Fermi statistics to the rotational states is not applied. Therefore, the spin statistics ({\it ortho}/{\it para}) does not appear in the rotational states of CH$_2$DCN and no correction is required.
For CH$_3$CN the 120$\degr$ and 240$\degr$ rotation can exactly interchange the two pairs of hydrogen nuclei. In this case, the total wave function must be symmetric with respect to the 120$\degr$ and 240$\degr$ rotation, according to the Fermi statistics. Considering that the $K$ states, except for $K$=0, are doubly degenerated (i.e. $\pm K$), the statistical weight of CH$_3$CN lines is 2 for $K$=3n (for n$\neq$0), and 1 for $K$=3n $\pm$ 1.
All the CH$_3$CN lines considered in our analysis have a statistical weight of 1, so no further correction is applied.
Finally, the line ratios are corrected for a factor (1-e$^{- \tau}$)/$\tau$ to account for the CH$_3$CN optical depths estimated from the LVG analysis described in Sect. \ref{subsec:nonLTE}.

\begin{figure}[h]
\begin{center}
\includegraphics[width=9cm]{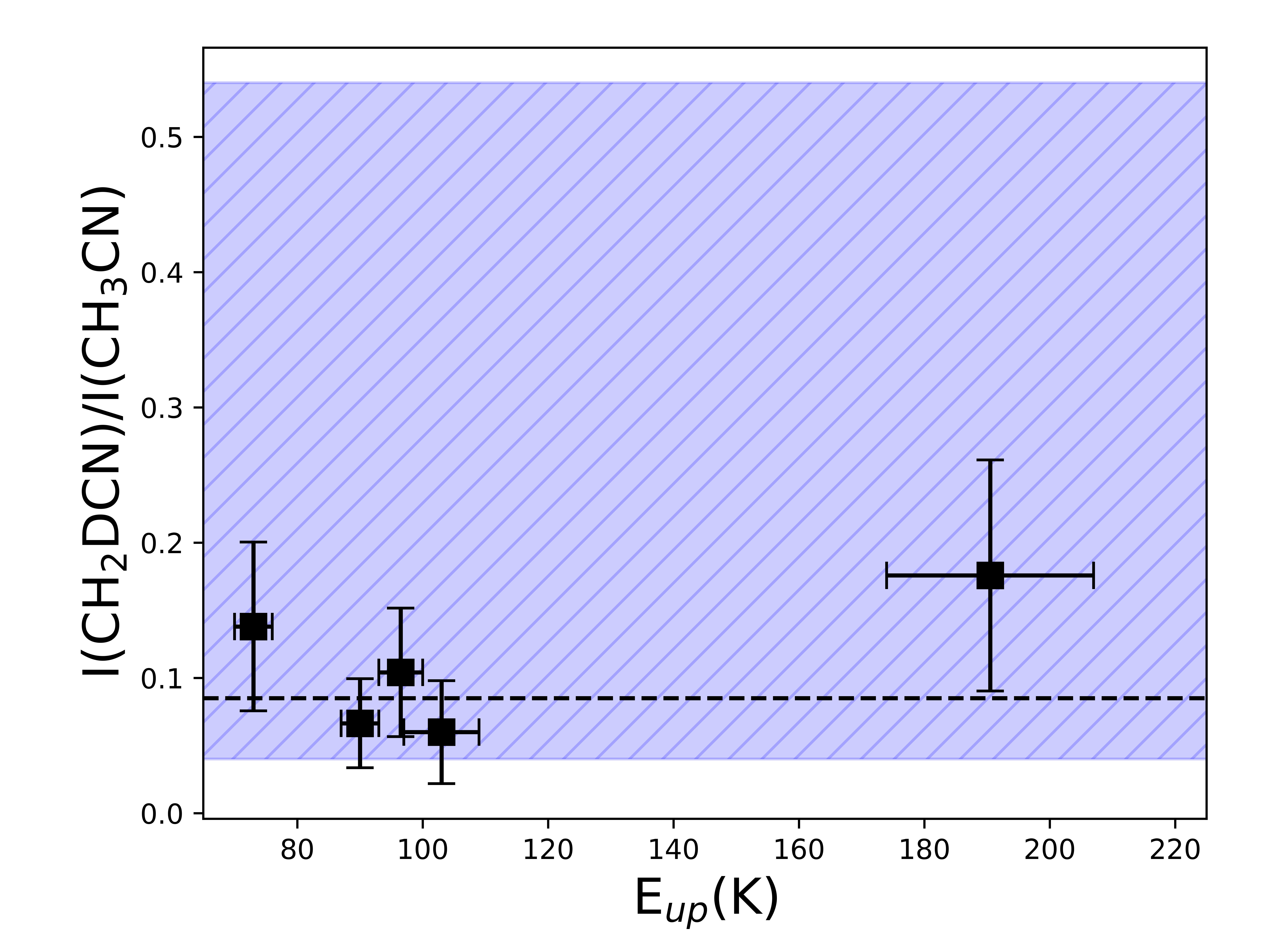}
  \caption{Deuteration of CH$_3$CN, derived using the line intensity ratios, as a function of the line upper-level energy E$_{\rm up}$. 
  The blue range indicates the deuteration derived using the N(CH$_2$DCN)/N(CH$_3$CN) column density ratio. 
  The dashed line indicates the best value of 9$\%$, which is consistent using the two methods described in Sect. \ref{subsec:CH2DCN-LTE}}
  \label{Fig:deut}
  \end{center}
\end{figure}

Figure \ref{Fig:deut} shows the [CH$_2$DCN]/[CH$_3$CN] derived with the method described above (i.e. from the intensity ratios) as a function of the upper level energy of the transition. 
The weighted average of the [CH$_2$DCN]/[CH$_3$CN] is $0.09\pm0.02$, consistent with the values derived by dividing the CH$_2$DCN and CH$_3$CN column densities (i.e. 4$\%$ -- 54$\%$), but with a smaller error bar (as expected).
Considering the presence of three H atoms, the enhancement of the elemental [D]/[H] is about 3$\%$.

CH$_3$CN has been recently detected towards SVS13-A also in the CALYPSO survey with the PdBI \citep{Belloche2020} and in the PEACHES survey with ALMA \citep{Yang2021}, even though with a lower number of detected lines (CALYPSO: 6; PEACHES: 3).
The CH$_3$CN column density derived by these two studies is perfectly consistent with the value derived in our analysis: $2 \times 10^{16}$ cm$^{-2}$ and $1 \times 10^{16}$ cm$^{-2}$, respectively, once scaled to the source size of 0$\farcs$3, derived by our non-LTE analysis. 
The PEACHES survey also reports  the detection of two CH$_2$DCN transitions and a column density of $3 \times 10^{15}$ cm$^{-2}$, which is very close to our value of $2 \times 10^{15}$ cm$^{-2}$.

\section{Discussion} \label{sec:discussion}
\begin{figure*}[h]
\begin{center}
\includegraphics[width=16cm]{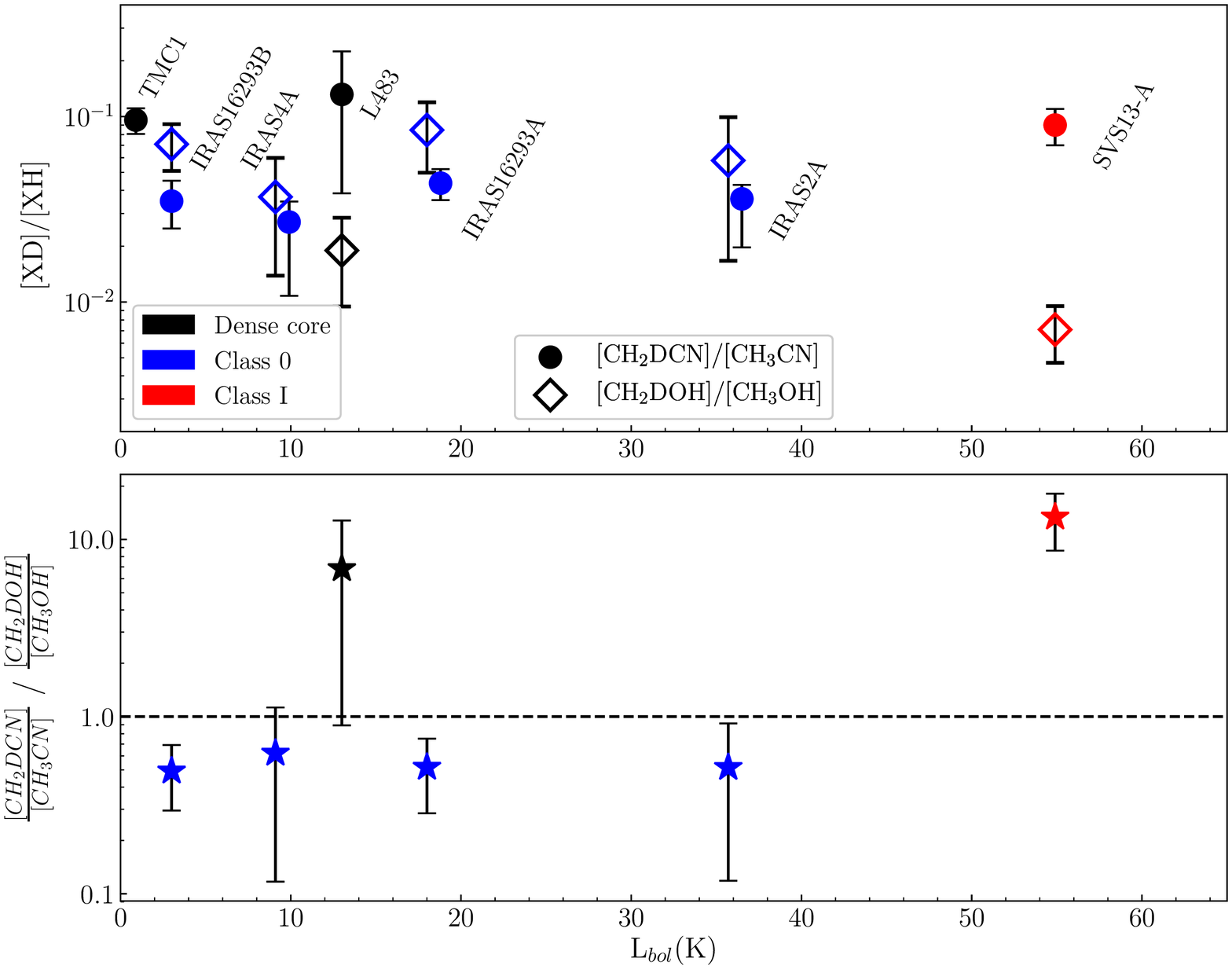}
  \caption{Comparison between SVS13-A, Class 0 sources and dense cores. {\it Upper Panel:} Deuteration of methyl cyanide (filled circles) and methanol (open diamonds), derived using the column density ratios of their deuterated and non-deuterated methyl group, as a function of the bolometric luminosity (L$_{bol}$) for dense cores (black symbols), Class 0 protostars (blue symbols), and the Class I protostar SVS13-A (red symbol). 
  The shown measurements are from \citet{Cabezas2021}, \citet{Agundez2019}, \citet{Calcutt2018}, \citet{Taquet2019},  \citet{Manigand2019}, \citet{Jorgensen2018}, \citet{Bianchi2017a}, and the present work (see text). L483 is an optical dark cloud core hosting a Class 0 protostar. However, the measurement by \citet{Agundez2019} refer to single-dish observations of the dense core around the protostar. The low rotational temperatures, the narrow FWHMs for the detected lines, and the high 30m beam dilution at 3mm further suggest that emission is arising mainly from the ambient cloud, and consequently it is classified as a dense core.
  Bolometric luminosities are from \citet{Kristensen2012}. {\it Lower Panel:} Ratio of methyl cyanide ([CH$_2$DCN]/[CH$_3$CN]) to methanol ([CH$_2$DOH]/[CH$_3$OH]) deuteration for each source, except TMC-1 for which methanol deuteration is not measured.}
  \label{Fig:deut-sources}
  \end{center}
\end{figure*}

\subsection{CH$_3$CN deuteration: Class 0 versus Class I hot corinos} \label{subsec:disc-ch3cn-class0classI}

While CH$_3$CN is very easily detected in young protostars, its deuterated isotopologue is not. 
We have detections of CH$_2$DCN in a handful of low-mass cold cores and protostars, so that the information about the degree of deuteration of this molecule is rather sparse.

To our best knowledge, in addition to except for SVS13-A, the deuteration of CH$_3$CN has been measured so far only towards the high-mass  star-forming region Sgr B2 \citep{Belloche2016}, and towards a limited number of Sun-like star-forming regions, namely: L483 \citep{Agundez2019}, TMC-1   \citep{Cabezas2021}, IRAS16293-2422 A and B \citep{Calcutt2018}, and NGC1333 IRAS 4A and IRAS 2A \citep{Taquet2019}. L483 is an optical dark cloud core hosting a Class 0 protostar. However, the measurement by \citet{Agundez2019} refers to single-dish observations of the dense core around the protostar. The low rotational temperatures, the narrow FWHMs for the detected lines, and the high IRAM-30m beam dilution at 3mm further suggest that emission is arising mainly from the ambient cloud and not from the Class 0 protostar.
In IRAS16293-2422 the [CH$_2$DCN]/[CH$_3$CN] abundances ratio is 4.4$\%$ for  protostar A and 3.5$\%$ for B. 
In NGC1333 IRAS 4A it is 2.7$\%$ and in IRAS 2A 3.6$\%$. 
In the L483 dense cold core it is 13$\%$ and in the cold core TMC-1 it is 9$\%$.
The situation is summarised in Fig. \ref{Fig:deut-sources} (upper panel).

The comparison between the above sources leads to two results.
First, cold cores seem to possess a higher deuteration degree than protostars.
Second, SVS13-A also seems   to have a CH$_3$CN deuteration higher by a factor of 2-3 with respect to Class 0 protostars.
We note that IRAS 2A and IRAS 4A are in the same star-forming region as SVS13-A, NGC 1333, so that in principle they experienced the same past thermal history.
In other words, if the CH$_3$CN deuteration was governed by the sublimation of the grain mantles, in principle there should not be a difference between these sources.

Finally, \cite{Yang2021} found a very tight correlation between methyl cyanide and methanol in the Class 0/I protostars of the Perseus molecular cloud to which SVS13-A belongs, which may imply a common origin of the two species.
Therefore, Fig. \ref{Fig:deut-sources} (upper panel) also shows the deuteration of methanol as measured on the methyl group, [CH$_2$DOH]/[CH$_3$OH], in the same sources where the [CH$_2$DCN]/[CH$_3$CN] was measured (with the exception of TMC-1 for which methanol deuteration is not measured).
Interestingly, the two values are approximately the same in Class 0 sources within the error bars, marginally different in the cold core L483, and different in SVS13-A.
Specifically, in SVS13-A, the [CH$_2$DCN]/[CH$_3$CN] ratio is about 14 times higher than the [CH$_2$DOH]/[CH$_3$OH] value. 
This would bring into question a possible common origin for the two species. However, methanol deuteration was   derived in SVS13-A using only single-dish observations \citep{Bianchi2017a}. High angular resolution interferometric observations are required to confirm this result.

\subsection{Chemistry of CH$_3$CN and CH$_2$DCN} \label{subsec:discussion-chemistry}

\begin{figure}
\begin{center}
\includegraphics[angle=0,width=9cm]{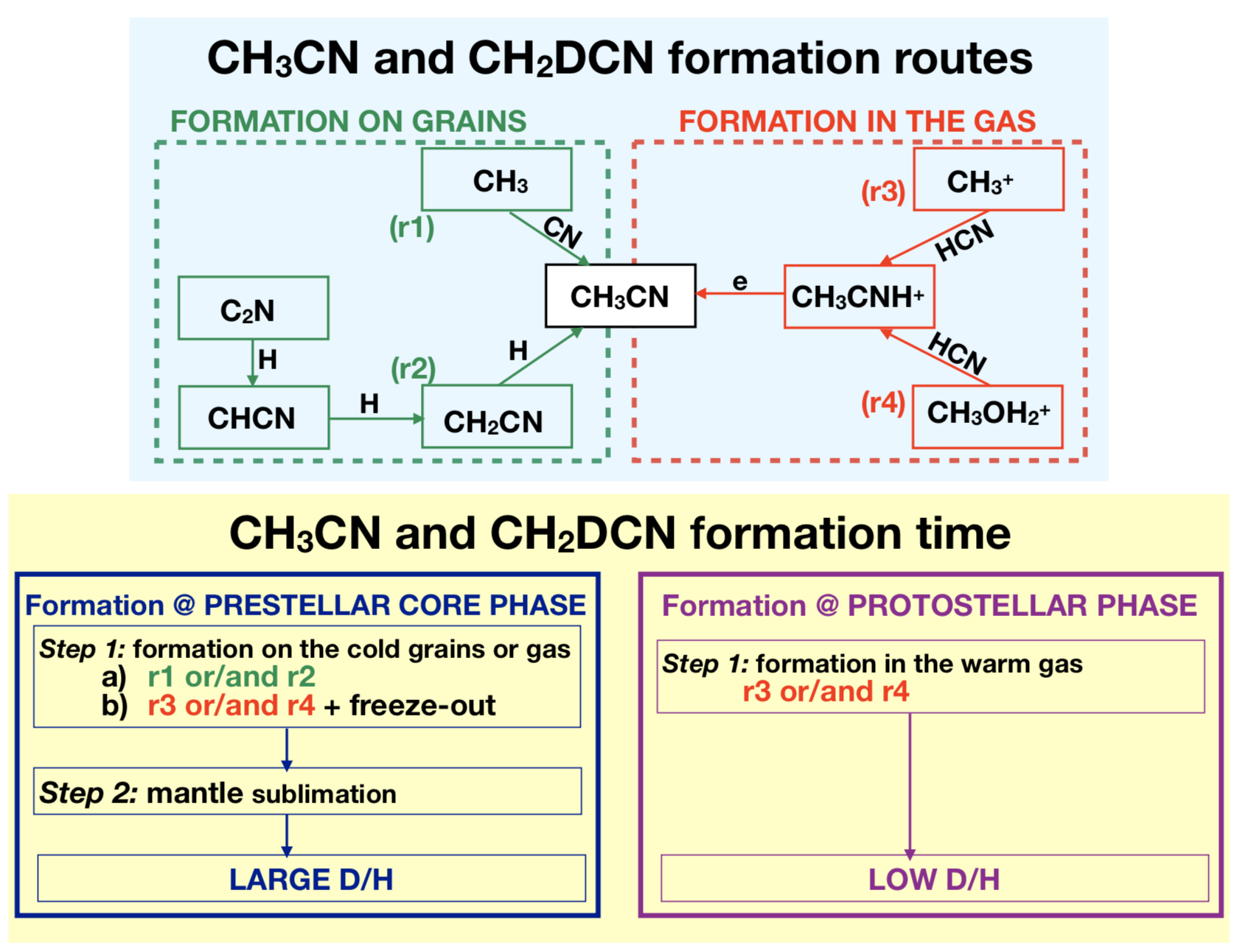} 
\caption{Scheme of the CH$_3$CN and CH$_2$DCN chemistry.
    \textit{Upper panel:} Different routes of formation on the grain surfaces (left) and in the gas phase (right).
    \textit{Lower panel:} Two possibilities for when CH$_3$CN and CH$_2$DCN formed, during the cold prestellar core phase (left) and during the protostellar warm phase (right). In the first case, high CH$_2$DCN/CH$_3$CN ratios ($\geq 0.1$) are expected, while in the second case the ratios are low ($\leq 0.1$).
}
\label{Fig:scheme-chemistry}
\end{center}
\end{figure}

In the literature two routes of methyl cyanide formation are invoked, either in the gas phase or on the surfaces of the grain mantles during the cold prestellar phase or in the warm protostellar phase.
Figure \ref{Fig:scheme-chemistry} provides a scheme of these various possibilities and their combination.
In the following we  review these possibilities and whether the measured CH$_3$CN deuteration can help to assess the dominant routes and the time of formation of CH$_3$CN and CH$_2$DCN.

\subsubsection{Formation routes of CH$_3$CN}

\textit{Formation in the gas:} 
As shown in the upper right panel of Fig. \ref{Fig:scheme-chemistry}, two possible routes are   invoked in the literature.
The first  involves the dissociative recombination of CH$_3$CNH$^+$ \citep{Vigren2008,Plessis2012}, which is in turn formed via the radiative association of CH$_3^+$ and HCN, whose rate constant is poorly known \citep[e.g.][and references therein]{Herbst1985,LeGal2019a}.
In addition, the reaction between protonated methanol and HCN is also a possible important route of CH$_3$CN formation \citep{Meot-Ner1986}, where abundant methanol and proton donors, such as H$_3^+$ or H$_3$O$^+$, are present.

\noindent
\textit{Formation on grains:} 
As shown in the upper left panel of Fig. \ref{Fig:scheme-chemistry}, methyl cyanide can be formed either by the combination of the two radicals CH$_3$ and CN or by the hydrogenation of C$_2$N \citep{Garrod2008}.
Unfortunately,  experimental or theoretical data are not available for either of these two routes, so their rate of formation in the current astrochemical models are estimated to have efficiency 1.
While this is certainly true for the C$_2$N hydrogenation, it is not clear that this is the case for the CH$_3$ and CN combination.
Ab initio theoretical studies have shown that the combination of two radicals on the icy grain surfaces can have barriers that reduce the efficiency of the reaction \citep[e.g.][]{Rimola2018,Enrique-Romero2019}.
Even though these authors did not explicitly study the CH$_3$CN case, their results caution on the assumption that radical-radical reactions always end up in iCOMs.

For completeness, we note that methanol, is believed to be synthesised on the grain surfaces by successive addition of hydrogen atoms \citep[e.g.][]{Watanabe2002,Rimola2014}.

\subsubsection{Destruction of  CH$_2$DCN in the gas phase}\label{subsec:destruction-ch2dcn}
In the gas phase, neutral species are predominantly destroyed by the most abundant molecular ions, such as H$_3^+$ or H$_2$DO$^+$, the latter where water is abundant, for example in warm regions.
Therefore, in the warm gas of Class 0/I protostars, the [CH$_2$DCN]/[CH$_3$CN] deuteration ratio tends to the values of [H$_2$D$^+$]/[H$_3^+$] and/or [H$_2$DO$^+$]/[H$_3$O$^+$].
Since H$_2$D$^+$ and H$_3^+$ are, by definition, formed in  warm gas, their [H$_2$D$^+$]/[H$_3^+$] abundance ratio is low \citep[e.g.][]{Charnley1997,Ceccarelli2014}.
The same applies to the protonated water, as it is mainly formed by the reaction of H$_3^+$ with H$_2$O, which is much less deuterated than CH$_3$CN \citep[e.g.][]{Coutens2012}.
The re-formation of CH$_3$CN and CH$_2$DCN in the gas phase via  reaction r4 shown in  Fig. \ref{Fig:scheme-chemistry}, will therefore tend to lower the [CH$_2$DCN]/[CH$_3$CN] abundance ratio.

On the contrary, the methanol major destruction route in warm gas is the reaction with OH \citep[e.g.][]{Shannon2013,codella2020}, which does not alter  methanol deuteration.

\subsubsection{Formation time}
Since  molecular deuteration is strongly impacted by  temperature, it is often used to disentangle whether a species is formed during the prestellar cold phase, frozen on the grain mantles, and then injected into the gas phase during the warm protostellar phase, or rather directly synthesised in the warm gas \citep[e.g.][]{Walmsley1989,Ceccarelli2007}, as illustrated in the lower panel of Fig. \ref{Fig:scheme-chemistry}.
In the first case, large deuteration factors are expected due to the low temperatures and CO depletion of the gas and dust, whereas the direct formation of molecules in warm gas leads to a much smaller deuteration factor \citep[see e.g.][]{Ceccarelli2014}.

The relatively high measured [CH$_2$DCN]/[CH$_3$CN] ratio (Fig. \ref{Fig:deut}) in the Class 0/I sources plays in favour of a formation of CH$_3$CN and CH$_2$DCN during the cold prestellar phase and their injection into the gas phase from the grain mantles once the protostar is formed.
Once CH$_3$CN and CH$_2$DCN are injected into the gas phase, reactions with H$_3^+$ will slowly decrease the [CH$_2$DCN]/[CH$_3$CN] ratio.
The possibly higher [CH$_2$DCN]/[CH$_3$CN] ratio in cold cores with respect to that measured in Class 0 protostars  perfectly agree with this hypothesis.
Likewise, the similar large deuteration of methanol in the same sources supports the formation of the two species during the cold prestellar phase.

Even so, it remains to be seen whether the sublimated CH$_3$CN and CH$_2$DCN, observed in cold dense cores and Class 0/I protostars, are grain-surfaces products or rather the result of the freezing-out of CH$_3$CN and CH$_2$DCN onto the grain mantles, as indicated in the left lower panel of Fig. \ref{Fig:scheme-chemistry}.
The two cases are   discussed separately in the following because, in principle, different routes could be dominant in the two classes of objects.

\noindent
\textit{Cold dense cores:}
The presence of gaseous CH$_3$CN in the cold cores favours the gas-phase formation hypothesis because   an additional process would be needed to extract methyl cyanide from the iced mantles at $\sim$10 K, a process that is not entirely clear.
Often,  the non-thermal desorption caused by the residual reaction energy not absorbed by the grains is invoked,  and called chemical desorption \citep[e.g.][]{Duley1993,Minissale2016}.
However, ab initio molecular dynamics computations on HCO challenge the idea  that a large fraction of the species formed on the grain icy surfaces can be released in the gas as the ices are very efficient in absorbing the reaction energy \citep{Pantaleone2020,Pantaleone2021}.
Finally, further support to the gas-phase synthesis in cold cores is provided by the modelling performed by \cite{Cabezas2021}, which claims that methyl cyanide and the measured [CH$_2$DCN]/[CH$_3$CN] ratio in TMC-1 are quite well reproduced by gas-phase formation routes.

\noindent
\textit{Class 0/I protostars:}
The situation in Class 0/I protostars is more complicated than that in the cold dense cores.
The similar deuteration of methyl cyanide and methanol and the tight correlation between CH$_3$OH and CH$_3$CN seen by \cite{Yang2021} in Class 0/I protostars would favour the hypothesis that either methyl cyanide is formed on the grain surfaces or the gas-phase reaction of HCN with protonated methanol is its major formation route.
However, since the deuteration during the cold prestellar core is governed by the enhancement of H$_2$D$^+$ with respect to H$_3^+$, regardless of the formation of the species on the grain surfaces (such as methanol) or in the gas phase (such as, possibly, CH$_3^+$ and HCN) \citep[e.g.][]{Ceccarelli2014}, the  similar deuteration degree of methyl cyanide and methanol cannot be used to discriminate between whether methyl cyanide is formed on the grains or in the gas.

Intriguingly, in SVS13-A both the [CH$_2$DCN]/[CH$_3$CN] and [CH$_2$DOH]/[CH$_3$OH] ratios are different from those of the Class 0 sources (Fig. \ref{Fig:deut-sources}): the first is about two to three times higher while the second is about ten times lower.
This would lead to thinking that either deuterated methyl cyanide and methanol are differently affected by gas-phase reactions or they were different already on the grain mantles.
As discussed in \S ~\ref{subsec:destruction-ch2dcn},  it is indeed possible that deuterated methyl cyanide and methanol are differently affected by gas-phase reactions.
However, methyl cyanide deuteration should diminish faster than that of methanol, contrarily to what we observe.

It seems  in SVS13-A, therefore, that methyl cyanide and methanol possess a different deuteration already on the grain surfaces.
Since the grain-surface synthesis of methyl cyanide involves CH$_3$ (+ CN), where CH$_3$ is a radical from the photolysis and/or radiolysis of methanol \citep[e.g.][]{Garrod2008}, the deuteration of methyl cyanide cannot differ from that of methanol if this is the major route.
A similar argument applies if CH$_3$CN is formed on the grain surfaces by the hydrogenation of C$_2$N.
It would then remain the possibility that methyl cyanide is more deuterated because it was formed during the prestellar phase in the gas phase by the reactions chain started by CH$_3^+$ (see above).
Since the deuteration from CH$_2$D$^+$ is active at higher temperatures than those where H$_2$D$^+$ is, this would explain the higher deuteration of CH$_3$CN with respect to CH$_3$OH in SVS13-A.

In summary, it seems likely that methyl cyanide in SVS13-A was synthesised in the gas phase of the cold prestellar phase and frozen out onto the grain mantles, from which it was injected into the gas phase again when the dust temperature reached the mantle's sublimation temperature.

\subsection{Structure of the SVS13-A hot corino} \label{subsec:molecules}

\begin{figure}
\begin{center}
\includegraphics[angle=0,width=10cm]{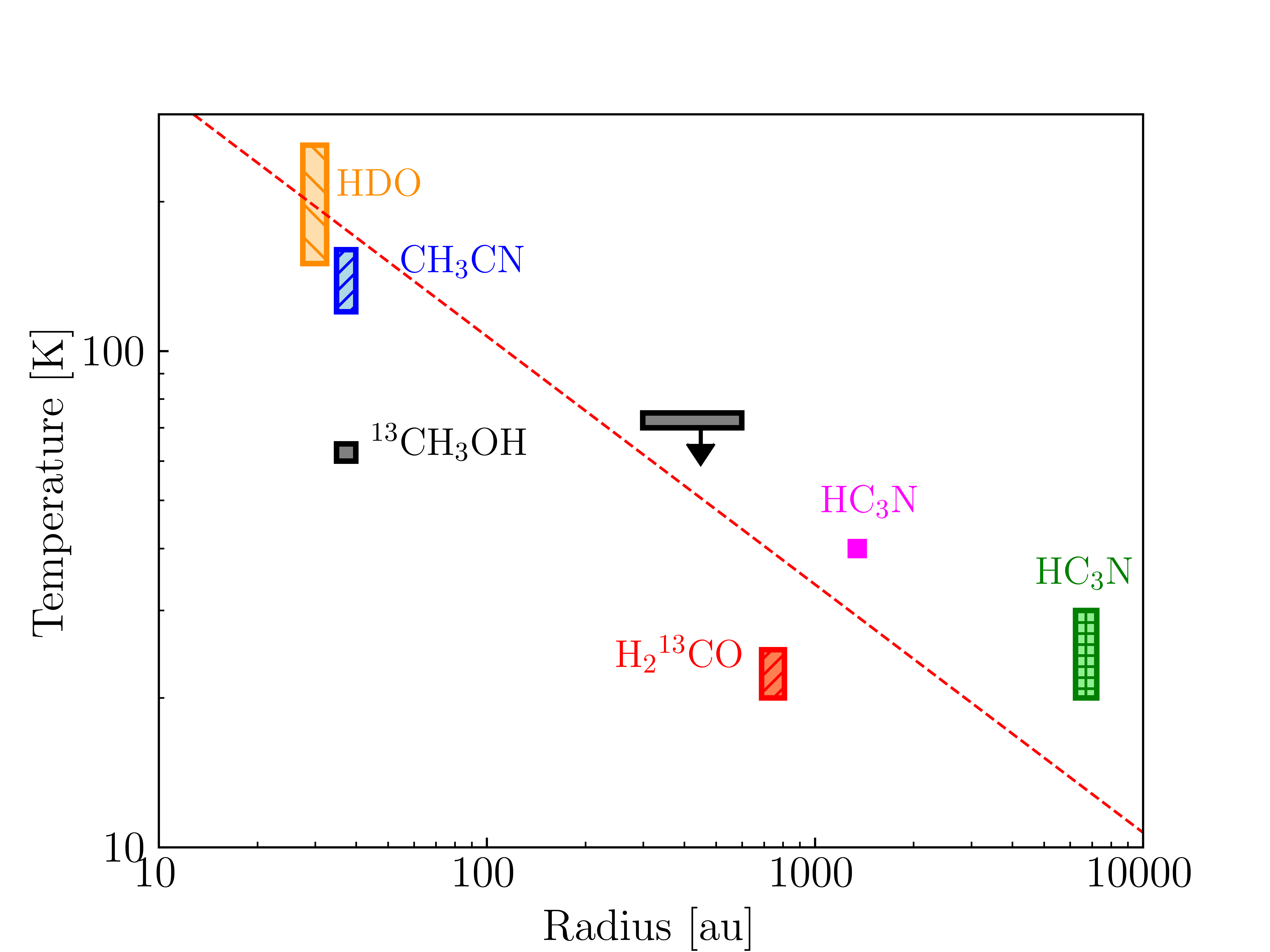} 
  \caption{Temperature profile of SVS13-A envelope as traced by different molecular species. The dashed red line is the theoretical temperature profile for a bolometric luminosity of 54.9 L$_{\odot}$ (from \citealt{Tobin2016}), and scaled to the NGC1333 distance of 299 pc obtained by
  {\it Gaia} \citep{Zucker2018}.} 
  \label{Fig:profiles}
  \end{center}
\end{figure}

SVS13-A is  one of the few young protostars for which an extensive analysis of different molecular tracers has been performed \citep{Lefloch1998a, Codella2016b, Bianchi2017a,Lefevre2017, Desimone2017, Bianchi2019a, Bianchi2019b}.
The non-LTE LVG analysis of formaldehyde, deuterated water, methanol, methyl cyanide, and cyanoacetylene offers an invaluable  opportunity to reconstruct its envelope temperature profile, as each species samples a different region of it.
Figure \ref{Fig:profiles} shows the derived gas temperature profile
as a function of the radius, as reconstructed putting together the various results.
In the same figure the dashed red line is the theoretical temperature profile for the bolometric luminosity of SVS13-A (L$_{bol}$ = 54.9 L$_{\odot}$ from \citealt{Tobin2016}, and scaled to the recently revised NGC1333 distance of 299 pc \citealt{Zucker2018}). 
The observational measurements are from \citet{Bianchi2017a}, \citet{Codella2016b}, \citet{Bianchi2019b}, and this paper.
CH$_3$CN traces the region where 
the abundances of (deuterated) water and methanol are
also enhanced due to the  hot corino activity
($\sim$ 0$\farcs$3, T $>$ 80 K).
Methanol also has  a warm component ($\leq$ 70 K) emitted in a region of $\sim$ 300-600 au in radius. 
For radii larger than 100 au, the molecular emission allows us to   reconstruct the temperature profiles and highlights the onion-like structure of the protostellar envelope \citep{Crimier2010, Jorgensen2002}. 
In particular, HC$_3$N traces one cold ($\sim$ 20 K) extended component, probably associated with the protostellar envelope, and a second lukewarm component ($\sim$ 40 K). H${_2}^{13}$CO is emitted from a region of $\sim$ 750 au in radius and it has a T$_{kin}$ between 20 and 25 K.  
In the inner 100 au the physical structure is indeed more complex, since the source is known to be a close binary system \citep{Anglada2000}. In summary, the chemical differentiation
as observed towards SVS13-A is consistent with a temperature
gradient due to protostellar heating.
Clearly, a proper modelling on scales of less than 50 au
will be done after sampling the chemical richness around
the two protostars of the SVS13-A system. 


\section{Conclusions} \label{sec:conclusions}

We analysed the CH$_3$CN and CH$_2$DCN emission in the Class I protostar SVS13-A, in the framework of the IRAM/NOEMA SOLIS and IRAM-30m ASAI Large Programs. Our conclusions can be summarised as follows: 

\begin{enumerate}

    \item[-] We detected 41 lines of CH$_3$CN and 7 lines of CH$_2$DCN, covering upper level energies (E$_{\rm up}$) from 13 K to 442 K and from 18 K to 200 K, respectively. The majority of the lines were detected using the IRAM-30m antenna, while two CH$_2$DCN lines were mapped using the IRAM/NOEMA interferometer.
    
    \item[-] The NOEMA maps show that the emission is concentrated towards SVS13-A and unresolved in a beam of 1$\farcs$5, consistent with the hypothesis that the CH$_2$DCN emission originates from the hot corino region.
    
     \item[-] We performed a non-LTE large velocity gradient analysis of the CH$_3$CN lines and derived a kinetic temperature of (140$\pm$20) K, a column density of (0.5-5) $\times$ 10$^{16}$ cm$^{-2}$, a gas density of n$_{\rm H_2}$ $\geq$ 10$^{7}$ cm$^{-3}$, and an emitting size of $\sim$0$\farcs$3. 
     Therefore, the non-LTE analysis confirms that  CH$_3$CN is also emitted from the hot corino region. The CH$_3$CN lines are predicted to be optically thick, with $\tau$ up to 5.
     
     \item[-] We performed a LTE rotation diagram analysis of the CH$_2$DCN lines assuming the same temperature (140 $\pm$ 20 K) and size (0$\farcs$3) derived by the CH$_3$CN analysis. 
     The CH$_2$DCN column density is $2.4^{+0.3}_{-0.4} \times 10^{15}$ cm$^{-2}$. The lines are predicted to be optically thin, with opacities lower than 0.2.
     
     \item[-] We derived CH$_3$CN deuteration, for the first time in a Class I protostar, using two different methods:   from the CH$_2$DCN/CH$_3$CN column density ratio and using the intensity ratio from lines with the same quantum number. 
     The first method gives an methyl cyanide deuteration between 0.04 and 0.54 with a best value of 0.1.
     The second method allows us to better constrain   the methyl cyanide deuteration because it does not depend on the gas temperature. 
     It yields a methyl cyanide deuteration CH$_2$DCN/CH$_3$CN equal to (0.09$\pm$0.02). 
     
     \item[-] The CH$_2$DCN/CH$_3$CN measured in SVS13-A is consistent with that observed in prestellar cores but it is a factor of  2--3 higher than the values observed in Class 0 protostars.
     We conclude that CH$_3$CN deuteration does not show a drastic decrease from the prestellar and Class 0 to the more evolved Class I phases.
     In addition, the methyl cyanide deuteration in IRAS 4A and IRAS 2A, two Class 0 protostars located in the same cloud as SVS13-A, NGC1333, is also about a factor of 2 lower than in SVS13-A. 
     This suggests that the SVS13-A greater methyl cyanide deuteration is not related to environmental conditions, such as the temperature in the NGC1333 region, at the epoch of ice formation. 
     
     \item[-] In SVS13-A, the CH$_3$CN deuteration is higher than for methanol, while they are approximately the same in Class 0 sources. 
     This seems to question a common origin for the two species. 
     We speculate that, in SVS13-A, methyl cyanide was synthesised in the gas phase by the reaction chain started from CH$_3^+$ in the cold prestellar phase, condensed onto the grain mantles, and was then injected back into the gas phase when the dust temperature reached the mantles sublimation temperature. 
     
     \item[-] Thanks to the analysis of different molecular tracers we reconstructed the source temperature profile, from the inner hot corino region to the extended envelope ($\sim$ 10000 au). 
     The temperature gradient is consistent with the SVS13-A bolometric luminosity of 55 L${\rm \odot}$. 
\end{enumerate}
  The physical structure of the inner regions will be further investigated by sampling the chemical complexity of the SVS13-A binary system on spatial scales smaller than 50 au.




\begin{acknowledgements}
While the paper was under the review process the detection of CH$_{\rm 2}$DCN has been also reported towards the source by \citet{Diaz2021}. Moreover, a measurement of CH$_{\rm 3}$CN deuteration in another Class I source (Ser-emb11) has been reported by \citet{Martin2021}.
We are very grateful to all the IRAM staff, whose dedication allowed us to carry out the SOLIS project. 
We are also grateful to Prof. Sonia Melandri for illuminating discussion on the spectroscopy of CH$_3$CN and CH$_2$DCN.
This project has received funding within the European Union’s Horizon 2020 research and innovation programme from the European Research Council (ERC) for the project “The Dawn of Organic Chemistry” (DOC), grant agreement No 741002, and from the Marie Sklodowska-Curie for the project ”Astro-Chemical Origins” (ACO), grant agreement No 811312.
This work was supported by the PRIN-INAF 2016 "The Cradle of Life - GENESIS-SKA (General Conditions in Early Planetary Systems for the rise of life with SKA)".

\end{acknowledgements}

\bibliographystyle{aa} 
\bibliography{SVS13A-CH2DCN-SOLIS} 

\begin{thebibliography}{89}
\expandafter\ifx\csname natexlab\endcsname\relax\def\natexlab#1{#1}\fi

\bibitem[{{Ag{\'u}ndez} {et~al.}(2019){Ag{\'u}ndez}, {Marcelino}, {Cernicharo},
  {Roueff}, \& {Tafalla}}]{Agundez2019}
{Ag{\'u}ndez}, M., {Marcelino}, N., {Cernicharo}, J., {Roueff}, E., \&
  {Tafalla}, M. 2019, \aap, 625, A147

\bibitem[{{Ag\'undez, M.} {et~al.}(2021){Ag\'undez, M.}, {Marcelino, N.},
  {Tercero, B.}, {Cabezas, C.}, {de Vicente, P.}, \& {Cernicharo,
  J.}}]{Agundez2021}
{Ag\'undez, M.}, {Marcelino, N.}, {Tercero, B.}, {et~al.} 2021, A\&A, 649, L4

\bibitem[{{Aikawa} {et~al.}(2012){Aikawa}, {Wakelam}, {Hersant}, {Garrod}, \&
  {Herbst}}]{Aikawa2012}
{Aikawa}, Y., {Wakelam}, V., {Hersant}, F., {Garrod}, R.~T., \& {Herbst}, E.
  2012, \apj, 760, 40

\bibitem[{{ALMA Partnership} {et~al.}(2015){ALMA Partnership}, {Brogan},
  {P{\'e}rez}, {Hunter}, {Dent}, {Hales}, {Hills}, {Corder}, {Fomalont},
  {Vlahakis}, {Asaki}, {Barkats}, {Hirota}, {Hodge}, {Impellizzeri}, {Kneissl},
  {Liuzzo}, {Lucas}, {Marcelino}, {Matsushita}, {Nakanishi}, {Phillips},
  {Richards}, {Toledo}, {Aladro}, {Broguiere}, {Cortes}, {Cortes}, {Espada},
  {Galarza}, {Garcia-Appadoo}, {Guzman-Ramirez}, {Humphreys}, {Jung}, {Kameno},
  {Laing}, {Leon}, {Marconi}, {Mignano}, {Nikolic}, {Nyman}, {Radiszcz},
  {Remijan}, {Rod{\'o}n}, {Sawada}, {Takahashi}, {Tilanus}, {Vila Vilaro},
  {Watson}, {Wiklind}, {Akiyama}, {Chapillon}, {de Gregorio-Monsalvo}, {Di
  Francesco}, {Gueth}, {Kawamura}, {Lee}, {Nguyen Luong}, {Mangum}, {Pietu},
  {Sanhueza}, {Saigo}, {Takakuwa}, {Ubach}, {van Kempen}, {Wootten},
  {Castro-Carrizo}, {Francke}, {Gallardo}, {Garcia}, {Gonzalez}, {Hill},
  {Kaminski}, {Kurono}, {Liu}, {Lopez}, {Morales}, {Plarre}, {Schieven},
  {Testi}, {Videla}, {Villard}, {Andreani}, {Hibbard}, \&
  {Tatematsu}}]{ALMA2015}
{ALMA Partnership}, {Brogan}, C.~L., {P{\'e}rez}, L.~M., {et~al.} 2015, \apjl,
  808, L3

\bibitem[{{Altwegg} {et~al.}(2019){Altwegg}, {Balsiger}, \&
  {Fuselier}}]{Altwegg2019}
{Altwegg}, K., {Balsiger}, H., \& {Fuselier}, S.~A. 2019, \araa, 57, 113

\bibitem[{{Andr{\'e}} {et~al.}(1993){Andr{\'e}}, {Ward-Thompson}, \&
  {Barsony}}]{Andre1993}
{Andr{\'e}}, P., {Ward-Thompson}, D., \& {Barsony}, M. 1993, \apj, 406, 122

\bibitem[{{Anglada} {et~al.}(2000){Anglada}, {Rodr{\'{\i}}guez}, \&
  {Torrelles}}]{Anglada2000}
{Anglada}, G., {Rodr{\'{\i}}guez}, L.~F., \& {Torrelles}, J.~M. 2000, \apjl,
  542, L123

\bibitem[{{Bachiller} {et~al.}(1998){Bachiller}, {Guilloteau}, {Gueth},
  {Tafalla}, {Dutrey}, {Codella}, \& {Castets}}]{Bachiller1998}
{Bachiller}, R., {Guilloteau}, S., {Gueth}, F., {et~al.} 1998, \aap, 339, L49

\bibitem[{{Belloche} {et~al.}(2020){Belloche}, {Maury}, {Maret}, {Anderl},
  {Bacmann}, {Andr{\'e}}, {Bontemps}, {Cabrit}, {Codella}, {Gaudel}, {Gueth},
  {Lef{\`e}vre}, {Lefloch}, {Podio}, \& {Testi}}]{Belloche2020}
{Belloche}, A., {Maury}, A.~J., {Maret}, S., {et~al.} 2020, \aap, 635, A198

\bibitem[{{Belloche} {et~al.}(2016){Belloche}, {M{\"u}ller}, {Garrod}, \&
  {Menten}}]{Belloche2016}
{Belloche}, A., {M{\"u}ller}, H.~S.~P., {Garrod}, R.~T., \& {Menten}, K.~M.
  2016, \aap, 587, A91

\bibitem[{{Bergner} {et~al.}(2018){Bergner}, {Guzm{\'a}n}, {{\"O}berg},
  {Loomis}, \& {Pegues}}]{Bergner2018}
{Bergner}, J.~B., {Guzm{\'a}n}, V.~G., {{\"O}berg}, K.~I., {Loomis}, R.~A., \&
  {Pegues}, J. 2018, \apj, 857, 69

\bibitem[{{Bianchi} {et~al.}(2019{\natexlab{a}}){Bianchi}, {Ceccarelli},
  {Codella}, {Enrique-Romero}, {Favre}, \& {Lefloch}}]{Bianchi2019b}
{Bianchi}, E., {Ceccarelli}, C., {Codella}, C., {et~al.} 2019{\natexlab{a}},
  ACS Earth and Space Chemistry, 3, 2659

\bibitem[{{Bianchi} {et~al.}(2017){Bianchi}, {Codella}, {Ceccarelli},
  {Fontani}, {Testi}, {Bachiller}, {Lefloch}, {Podio}, \&
  {Taquet}}]{Bianchi2017a}
{Bianchi}, E., {Codella}, C., {Ceccarelli}, C., {et~al.} 2017, \mnras, 467,
  3011

\bibitem[{{Bianchi} {et~al.}(2019{\natexlab{b}}){Bianchi}, {Codella},
  {Ceccarelli}, {Vazart}, {Bachiller}, {Balucani}, {Bouvier}, {De Simone},
  {Enrique-Romero}, {Kahane}, {Lefloch}, {L{\'o}pez-Sepulcre},
  {Ospina-Zamudio}, {Podio}, \& {Taquet}}]{Bianchi2019a}
{Bianchi}, E., {Codella}, C., {Ceccarelli}, C., {et~al.} 2019{\natexlab{b}},
  \mnras, 483, 1850

\bibitem[{{Cabezas} {et~al.}(2021){Cabezas}, {Endo, Y.}, {Roueff, E.},
  {Marcelino, N.}, {Ag\'undez, M.}, {Tercero, B.}, \& {Cernicharo,
  J.}}]{Cabezas2021}
{Cabezas}, C., {Endo, Y.}, {Roueff, E.}, {et~al.} 2021, A\&A, 646, L1

\bibitem[{{Calcutt} {et~al.}(2018){Calcutt}, {J{\o}rgensen}, {M{\"u}ller},
  {Kristensen}, {Coutens}, {Bourke}, {Garrod}, {Persson}, {van der Wiel}, {van
  Dishoeck}, \& {Wampfler}}]{Calcutt2018}
{Calcutt}, H., {J{\o}rgensen}, J.~K., {M{\"u}ller}, H.~S.~P., {et~al.} 2018,
  \aap, 616, A90

\bibitem[{{Caselli} \& {Ceccarelli}(2012)}]{Caselli2012}
{Caselli}, P. \& {Ceccarelli}, C. 2012, \aapr, 20, 56

\bibitem[{{Cazzoli} \& {Puzzarini}(2006)}]{Cazzoli2006}
{Cazzoli}, G. \& {Puzzarini}, C. 2006, Journal of Molecular Spectroscopy, 240,
  153

\bibitem[{{Ceccarelli} {et~al.}(2014){Ceccarelli}, {Caselli},
  {Bockel{\'e}e-Morvan}, {Mousis}, {Pizzarello}, {Robert}, \&
  {Semenov}}]{Ceccarelli2014}
{Ceccarelli}, C., {Caselli}, P., {Bockel{\'e}e-Morvan}, D., {et~al.} 2014,
  Protostars and Planets VI, 859

\bibitem[{{Ceccarelli} {et~al.}(2017){Ceccarelli}, {Caselli}, {Fontani},
  {Neri}, {L{\'o}pez-Sepulcre}, {Codella}, {Feng}, {Jim{\'e}nez-Serra},
  {Lefloch}, {Pineda}, {Vastel}, {Alves}, {Bachiller}, {Balucani}, {Bianchi},
  {Bizzocchi}, {Bottinelli}, {Caux}, {Chac{\'o}n-Tanarro}, {Choudhury},
  {Coutens}, {Dulieu}, {Favre}, {Hily-Blant}, {Holdship}, {Kahane}, {Jaber
  Al-Edhari}, {Laas}, {Ospina}, {Oya}, {Podio}, {Pon}, {Punanova}, {Quenard},
  {Rimola}, {Sakai}, {Sims}, {Spezzano}, {Taquet}, {Testi}, {Theul{\'e}},
  {Ugliengo}, {Vasyunin}, {Viti}, {Wiesenfeld}, \& {Yamamoto}}]{Ceccarelli2017}
{Ceccarelli}, C., {Caselli}, P., {Fontani}, F., {et~al.} 2017, \apj, 850, 176

\bibitem[{{Ceccarelli} {et~al.}(2007){Ceccarelli}, {Caselli}, {Herbst},
  {Tielens}, \& {Caux}}]{Ceccarelli2007}
{Ceccarelli}, C., {Caselli}, P., {Herbst}, E., {Tielens}, A.~G.~G.~M., \&
  {Caux}, E. 2007, Protostars and Planets V, 47

\bibitem[{{Ceccarelli} {et~al.}(2003){Ceccarelli}, {Maret}, {Tielens},
  {Castets}, \& {Caux}}]{Ceccarelli2003}
{Ceccarelli}, C., {Maret}, S., {Tielens}, A.~G.~G.~M., {Castets}, A., \&
  {Caux}, E. 2003, \aap, 410, 587

\bibitem[{{Charnley} {et~al.}(1997){Charnley}, {Tielens}, \&
  {Rodgers}}]{Charnley1997}
{Charnley}, S.~B., {Tielens}, A.~G.~G.~M., \& {Rodgers}, S.~D. 1997, \apjl,
  482, L203

\bibitem[{{Chen} {et~al.}(2009){Chen}, {Launhardt}, \& {Henning}}]{Chen2009}
{Chen}, X., {Launhardt}, R., \& {Henning}, T. 2009, \apj, 691, 1729

\bibitem[{{Chini} {et~al.}(1997){Chini}, {Reipurth}, {Sievers},
  {Ward-Thompson}, {Haslam}, {Kreysa}, \& {Lemke}}]{Chini1997}
{Chini}, R., {Reipurth}, B., {Sievers}, A., {et~al.} 1997, \aap, 325, 542

\bibitem[{{Codella} {et~al.}(1999){Codella}, {Bachiller}, \&
  {Reipurth}}]{Codella1999}
{Codella}, C., {Bachiller}, R., \& {Reipurth}, B. 1999, \aap, 343, 585

\bibitem[{{Codella} {et~al.}(2009){Codella}, {Benedettini}, {Beltr{\'a}n},
  {Gueth}, {Viti}, {Bachiller}, {Tafalla}, {Cabrit}, {Fuente}, \&
  {Lefloch}}]{Codella2009}
{Codella}, C., {Benedettini}, M., {Beltr{\'a}n}, M.~T., {et~al.} 2009, \aap,
  507, L25

\bibitem[{{Codella} {et~al.}(2020){Codella}, {Ceccarelli}, {Bianchi},
  {Balucani}, {Podio}, {Caselli}, {Feng}, {Lefloch}, {L{\'o}pez-Sepulcre},
  {Neri}, {Spezzano}, \& {De Simone}}]{codella2020}
{Codella}, C., {Ceccarelli}, C., {Bianchi}, E., {et~al.} 2020, \aap, 635, A17

\bibitem[{{Codella} {et~al.}(2016){Codella}, {Ceccarelli}, {Bianchi}, {Podio},
  {Bachiller}, {Lefloch}, {Fontani}, {Taquet}, \& {Testi}}]{Codella2016b}
{Codella}, C., {Ceccarelli}, C., {Bianchi}, E., {et~al.} 2016, \mnras, 462, L75

\bibitem[{{Codella} {et~al.}(2012){Codella}, {Ceccarelli}, {Lefloch},
  {Fontani}, {Busquet}, {Caselli}, {Kahane}, {Lis}, {Taquet}, {Vasta}, {Viti},
  \& {Wiesenfeld}}]{Codella2012}
{Codella}, C., {Ceccarelli}, C., {Lefloch}, B., {et~al.} 2012, \apjl, 757, L9

\bibitem[{{Coutens} {et~al.}(2016){Coutens}, {J{\o}rgensen}, {van der Wiel},
  {M{\"u}ller}, {Lykke}, {Bjerkeli}, {Bourke}, {Calcutt}, {Drozdovskaya},
  {Favre}, {Fayolle}, {Garrod}, {Jacobsen}, {Ligterink}, {{\"O}berg},
  {Persson}, {van Dishoeck}, \& {Wampfler}}]{Coutens2016}
{Coutens}, A., {J{\o}rgensen}, J.~K., {van der Wiel}, M.~H.~D., {et~al.} 2016,
  \aap, 590, L6

\bibitem[{{Coutens} {et~al.}(2012){Coutens}, {Vastel}, {Caux}, {Ceccarelli},
  {Bottinelli}, {Wiesenfeld}, {Faure}, {Scribano}, \& {Kahane}}]{Coutens2012}
{Coutens}, A., {Vastel}, C., {Caux}, E., {et~al.} 2012, \aap, 539, A132

\bibitem[{{Crimier} {et~al.}(2010){Crimier}, {Ceccarelli}, {Maret},
  {Bottinelli}, {Caux}, {Kahane}, {Lis}, \& {Olofsson}}]{Crimier2010}
{Crimier}, N., {Ceccarelli}, C., {Maret}, S., {et~al.} 2010, \aap, 519, A65

\bibitem[{{De Simone} {et~al.}(2017){De Simone}, {Codella}, {Testi},
  {Belloche}, {Maury}, {Anderl}, {Andr{\'e}}, {Maret}, \&
  {Podio}}]{Desimone2017}
{De Simone}, M., {Codella}, C., {Testi}, L., {et~al.} 2017, \aap, 599, A121

\bibitem[{{Diaz-Rodriguez} {et~al.}(2021){Diaz-Rodriguez}, {Anglada},
  {Bl{\'a}zquez-Calero}, {Osorio}, {G{\'o}mez}, {Fuller}, {Estalella},
  {Torrelles}, {Cabrit}, {Rodr{\'\i}guez}, {Lef{\`e}vre}, {Mac{\'\i}as},
  {Carrasco-Gonz{\'a}lez}, {Zapata}, {de Gregorio-Monsalvo}, \&
  {Ho}}]{Diaz2021}
{Diaz-Rodriguez}, A.~K., {Anglada}, G., {Bl{\'a}zquez-Calero}, G., {et~al.}
  2021, arXiv e-prints, arXiv:2111.11787

\bibitem[{{Duley} \& {Williams}(1993)}]{Duley1993}
{Duley}, W.~W. \& {Williams}, D.~A. 1993, \mnras, 260, 37

\bibitem[{{Enrique-Romero} {et~al.}(2019){Enrique-Romero}, {Rimola},
  {Ceccarelli}, {Ugliengo}, {Balucani}, \& {Skouteris}}]{Enrique-Romero2019}
{Enrique-Romero}, J., {Rimola}, A., {Ceccarelli}, C., {et~al.} 2019, ACS Earth
  and Space Chemistry, 3, 2158

\bibitem[{{Fedele} {et~al.}(2018){Fedele}, {Tazzari}, {Booth}, {Testi},
  {Clarke}, {Pascucci}, {Kospal}, {Semenov}, {Bruderer}, {Henning}, \&
  {Teague}}]{Fedele2018}
{Fedele}, D., {Tazzari}, M., {Booth}, R., {et~al.} 2018, \aap, 610, A24

\bibitem[{{Furuya} {et~al.}(2017){Furuya}, {Drozdovskaya}, {Visser}, {van
  Dishoeck}, {Walsh}, {Harsono}, {Hincelin}, \& {Taquet}}]{Furuya2017}
{Furuya}, K., {Drozdovskaya}, M.~N., {Visser}, R., {et~al.} 2017, \aap, 599,
  A40

\bibitem[{{Garrod} {et~al.}(2008){Garrod}, {Widicus Weaver}, \&
  {Herbst}}]{Garrod2008}
{Garrod}, R.~T., {Widicus Weaver}, S.~L., \& {Herbst}, E. 2008, \apj, 682, 283

\bibitem[{{Green}(1986)}]{Green1986}
{Green}, S. 1986, \apj, 309, 331

\bibitem[{{Herbst}(1985)}]{Herbst1985}
{Herbst}, E. 1985, \apj, 291, 226

\bibitem[{{Herbst} \& {van Dishoeck}(2009)}]{Herbst2009}
{Herbst}, E. \& {van Dishoeck}, E.~F. 2009, \araa, 47, 427

\bibitem[{{Jensen} {et~al.}(2021){Jensen}, {J\o{}rgensen}, {Kristensen, L. E.},
  {Coutens, A.}, {van Dishoeck, E. F.}, {Furuya, K.}, {Harsono, D.}, \&
  {Persson, M. V.}}]{Jensen2021}
{Jensen}, S.~S., {J\o{}rgensen}, J.~K., {Kristensen, L. E.}, {et~al.} 2021,
  A\&A, 650, A172

\bibitem[{{J{\o}rgensen} {et~al.}(2018){J{\o}rgensen}, {M{\"u}ller}, {Calcutt},
  {Coutens}, {Drozdovskaya}, {{\"O}berg}, {Persson}, {Taquet}, {van Dishoeck},
  \& {Wampfler}}]{Jorgensen2018}
{J{\o}rgensen}, J.~K., {M{\"u}ller}, H.~S.~P., {Calcutt}, H., {et~al.} 2018,
  \aap, 620, A170

\bibitem[{{J{\o}rgensen} {et~al.}(2002){J{\o}rgensen}, {Sch{\"o}ier}, \& {van
  Dishoeck}}]{Jorgensen2002}
{J{\o}rgensen}, J.~K., {Sch{\"o}ier}, F.~L., \& {van Dishoeck}, E.~F. 2002,
  \aap, 389, 908

\bibitem[{{Kahane} {et~al.}(2018){Kahane}, {Jaber Al-Edhari}, {Ceccarelli},
  {L{\'o}pez-Sepulcre}, {Fontani}, \& {Kama}}]{Kahane2018}
{Kahane}, C., {Jaber Al-Edhari}, A., {Ceccarelli}, C., {et~al.} 2018, \apj,
  852, 130

\bibitem[{{Kristensen} {et~al.}(2012){Kristensen}, {van Dishoeck}, {Bergin},
  {Visser}, {Y{\i}ld{\i}z}, {San Jose-Garcia}, {J{\o}rgensen}, {Herczeg},
  {Johnstone}, {Wampfler}, {Benz}, {Bruderer}, {Cabrit}, {Caselli}, {Doty},
  {Harsono}, {Herpin}, {Hogerheijde}, {Karska}, {van Kempen}, {Liseau},
  {Nisini}, {Tafalla}, {van der Tak}, \& {Wyrowski}}]{Kristensen2012}
{Kristensen}, L.~E., {van Dishoeck}, E.~F., {Bergin}, E.~A., {et~al.} 2012,
  \aap, 542, A8

\bibitem[{{Le Gal} {et~al.}(2019){Le Gal}, {Brady}, {{\"O}berg}, {Roueff}, \&
  {Le Petit}}]{LeGal2019a}
{Le Gal}, R., {Brady}, M.~T., {{\"O}berg}, K.~I., {Roueff}, E., \& {Le Petit},
  F. 2019, \apj, 886, 86

\bibitem[{{Le Gal} {et~al.}(2020){Le Gal}, {{\"O}berg}, {Huang}, {Law},
  {M{\'e}nard}, {Lefloch}, {Vastel}, {Lopez-Sepulcre}, {Favre}, {Bianchi}, \&
  {Ceccarelli}}]{LeGal2020}
{Le Gal}, R., {{\"O}berg}, K.~I., {Huang}, J., {et~al.} 2020, \apj, 898, 131

\bibitem[{{Le Roy} {et~al.}(2015){Le Roy}, {Altwegg}, {Balsiger}, {Berthelier},
  {Bieler}, {Briois}, {Calmonte}, {Combi}, {De Keyser}, {Dhooghe}, {Fiethe},
  {Fuselier}, {Gasc}, {Gombosi}, {H{\"a}ssig}, {J{\"a}ckel}, {Rubin}, \&
  {Tzou}}]{LeRoy2015}
{Le Roy}, L., {Altwegg}, K., {Balsiger}, H., {et~al.} 2015, \aap, 583, A1

\bibitem[{{Lef{\`e}vre} {et~al.}(2017){Lef{\`e}vre}, {Cabrit}, {Maury},
  {Gueth}, {Tabone}, {Podio}, {Belloche}, {Codella}, {Maret}, {Anderl},
  {Andr{\'e}}, \& {Hennebelle}}]{Lefevre2017}
{Lef{\`e}vre}, C., {Cabrit}, S., {Maury}, A.~J., {et~al.} 2017, \aap, 604, L1

\bibitem[{{Lefloch} {et~al.}(2018){Lefloch}, {Bachiller}, {Ceccarelli},
  {Cernicharo}, {Codella}, {Fuente}, {Kahane}, {L{\'o}pez-Sepulcre}, {Tafalla},
  {Vastel}, {Caux}, {Gonz{\'a}lez-Garc{\'{\i}}a}, {Bianchi}, {G{\'o}mez-Ruiz},
  {Holdship}, {Mendoza}, {Ospina-Zamudio}, {Podio}, {Qu{\'e}nard}, {Roueff},
  {Sakai}, {Viti}, {Yamamoto}, {Yoshida}, {Favre}, {Monfredini},
  {Quiti{\'a}n-Lara}, {Marcelino}, {Boechat-Roberty}, \&
  {Cabrit}}]{Lefloch2018}
{Lefloch}, B., {Bachiller}, R., {Ceccarelli}, C., {et~al.} 2018, \mnras, 477,
  4792

\bibitem[{{Lefloch} {et~al.}(1998){Lefloch}, {Castets}, {Cernicharo}, {Langer},
  \& {Zylka}}]{Lefloch1998a}
{Lefloch}, B., {Castets}, A., {Cernicharo}, J., {Langer}, W.~D., \& {Zylka}, R.
  1998, \aap, 334, 269

\bibitem[{{Loomis} {et~al.}(2018){Loomis}, {Cleeves}, {{\"O}berg}, {Aikawa},
  {Bergner}, {Furuya}, {Guzman}, \& {Walsh}}]{Loomis2018}
{Loomis}, R.~A., {Cleeves}, L.~I., {{\"O}berg}, K.~I., {et~al.} 2018, \apj,
  859, 131

\bibitem[{{Looney} {et~al.}(2000){Looney}, {Mundy}, \& {Welch}}]{Looney2000}
{Looney}, L.~W., {Mundy}, L.~G., \& {Welch}, W.~J. 2000, \apj, 529, 477

\bibitem[{{Manigand} {et~al.}(2019){Manigand}, {Calcutt}, {J{\o}rgensen},
  {Taquet}, {M{\"u}ller}, {Coutens}, {Wampfler}, {Ligterink}, {Drozdovskaya},
  {Kristensen}, {van der Wiel}, \& {Bourke}}]{Manigand2019}
{Manigand}, S., {Calcutt}, H., {J{\o}rgensen}, J.~K., {et~al.} 2019, \aap, 623,
  A69

\bibitem[{{Maret} {et~al.}(2011){Maret}, {Hily-Blant}, {Pety}, {Bardeau}, \&
  {Reynier}}]{Maret2011}
{Maret}, S., {Hily-Blant}, P., {Pety}, J., {Bardeau}, S., \& {Reynier}, E.
  2011, \aap, 526, A47

\bibitem[{{Martin-Domenech} {et~al.}(2021){Martin-Domenech}, {Bergner},
  {Oberg}, {Carpenter}, {Law}, {Huang}, {Jorgensen}, {Schwarz}, \&
  {Wilner}}]{Martin2021}
{Martin-Domenech}, R., {Bergner}, J.~B., {Oberg}, K.~I., {et~al.} 2021, arXiv
  e-prints, arXiv:2109.11512

\bibitem[{{Maury} {et~al.}(2019){Maury}, {Andr{\'e}}, {Testi}, {Maret},
  {Belloche}, {Hennebelle}, {Cabrit}, {Codella}, {Gueth}, {Podio}, {Anderl},
  {Bacmann}, {Bontemps}, {Gaudel}, {Ladjelate}, {Lef{\`e}vre}, {Tabone}, \&
  {Lefloch}}]{Maury2019}
{Maury}, A.~J., {Andr{\'e}}, P., {Testi}, L., {et~al.} 2019, \aap, 621, A76

\bibitem[{Meot-Ner \& Karpas(1986)}]{Meot-Ner1986}
Meot-Ner, M. \& Karpas, Z. 1986, The Journal of Physical Chemistry, 90, 2206

\bibitem[{{Minissale} {et~al.}(2016){Minissale}, {Dulieu}, {Cazaux}, \&
  {Hocuk}}]{Minissale2016}
{Minissale}, M., {Dulieu}, F., {Cazaux}, S., \& {Hocuk}, S. 2016, \aap, 585,
  A24

\bibitem[{{M{\"u}ller} {et~al.}(2009){M{\"u}ller}, {Drouin}, \&
  {Pearson}}]{Muller2009}
{M{\"u}ller}, H.~S.~P., {Drouin}, B.~J., \& {Pearson}, J.~C. 2009, \aap, 506,
  1487

\bibitem[{{M{\"u}ller} {et~al.}(2005){M{\"u}ller}, {Schl{\"o}der}, {Stutzki},
  \& {Winnewisser}}]{Muller2005}
{M{\"u}ller}, H.~S.~P., {Schl{\"o}der}, F., {Stutzki}, J., \& {Winnewisser}, G.
  2005, Journal of Molecular Structure, 742, 215

\bibitem[{{Nazari} {et~al.}(2021){Nazari}, {van Gelder}, {van Dishoeck},
  {Tabone}, {van't Hoff}, {Ligterink}, {Beuther}, {Boogert}, {Caratti o
  Garatti}, {Klaassen}, {Linnartz}, {Taquet}, \& {Tychoniec}}]{Nazari2021}
{Nazari}, P., {van Gelder}, M.~L., {van Dishoeck}, E.~F., {et~al.} 2021, \aap,
  650, A150

\bibitem[{{{\"O}berg} {et~al.}(2015){{\"O}berg}, {Guzm{\'a}n}, {Furuya}, {Qi},
  {Aikawa}, {Andrews}, {Loomis}, \& {Wilner}}]{Oberg2015}
{{\"O}berg}, K.~I., {Guzm{\'a}n}, V.~V., {Furuya}, K., {et~al.} 2015, \nat,
  520, 198

\bibitem[{{{\"O}berg} {et~al.}(2014){{\"O}berg}, {Lauck}, \&
  {Graninger}}]{Oberg2014}
{{\"O}berg}, K.~I., {Lauck}, T., \& {Graninger}, D. 2014, \apj, 788, 68

\bibitem[{{Pantaleone} {et~al.}(2021){Pantaleone}, {Enrique-Romero},
  {Ceccarelli}, {Ferrero}, {Balucani}, {Rimola}, \&
  {Ugliengo}}]{Pantaleone2021}
{Pantaleone}, S., {Enrique-Romero}, J., {Ceccarelli}, C., {et~al.} 2021, \apj,
  917, 49

\bibitem[{{Pantaleone} {et~al.}(2020){Pantaleone}, {Enrique-Romero},
  {Ceccarelli}, {Ugliengo}, {Balucani}, \& {Rimola}}]{Pantaleone2020}
{Pantaleone}, S., {Enrique-Romero}, J., {Ceccarelli}, C., {et~al.} 2020, \apj,
  897, 56

\bibitem[{{Pickett} {et~al.}(1998){Pickett}, {Poynter}, {Cohen}, {Delitsky},
  {Pearson}, \& {M{\"u}ller}}]{Pickett1998}
{Pickett}, H.~M., {Poynter}, R.~L., {Cohen}, E.~A., {et~al.} 1998, \jqsrt, 60,
  883

\bibitem[{{Plessis} {et~al.}(2012){Plessis}, {Carrasco}, {Dobrijevic}, \&
  {Pernot}}]{Plessis2012}
{Plessis}, S., {Carrasco}, N., {Dobrijevic}, M., \& {Pernot}, P. 2012, \icarus,
  219, 254

\bibitem[{{Reipurth} {et~al.}(1993){Reipurth}, {Chini}, {Krugel}, {Kreysa}, \&
  {Sievers}}]{Reipurth1993}
{Reipurth}, B., {Chini}, R., {Krugel}, E., {Kreysa}, E., \& {Sievers}, A. 1993,
  \aap, 273, 221

\bibitem[{{Rimola} {et~al.}(2018){Rimola}, {Skouteris}, {Balucani},
  {Ceccarelli}, {Enrique-Romero}, {Taquet}, \& {Ugliengo}}]{Rimola2018}
{Rimola}, A., {Skouteris}, D., {Balucani}, N., {et~al.} 2018, ACS Earth and
  Space Chemistry, 2, 720

\bibitem[{{Rimola} {et~al.}(2014){Rimola}, {Taquet}, {Ugliengo}, {Balucani}, \&
  {Ceccarelli}}]{Rimola2014}
{Rimola}, A., {Taquet}, V., {Ugliengo}, P., {Balucani}, N., \& {Ceccarelli}, C.
  2014, \aap, 572, A70

\bibitem[{{Sch{\"o}ier} {et~al.}(2005){Sch{\"o}ier}, {van der Tak}, {van
  Dishoeck}, \& {Black}}]{Schoier2005}
{Sch{\"o}ier}, F.~L., {van der Tak}, F.~F.~S., {van Dishoeck}, E.~F., \&
  {Black}, J.~H. 2005, \aap, 432, 369

\bibitem[{{Scoville} \& {Solomon}(1974)}]{Scoville1974}
{Scoville}, N.~Z. \& {Solomon}, P.~M. 1974, \apjl, 187, L67

\bibitem[{{Segura-Cox} {et~al.}(2020){Segura-Cox}, {Schmiedeke}, {Pineda},
  {Stephens}, {Fern{\'a}ndez-L{\'o}pez}, {Looney}, {Caselli}, {Li}, {Mundy},
  {Kwon}, \& {Harris}}]{Segura-Cox2020}
{Segura-Cox}, D.~M., {Schmiedeke}, A., {Pineda}, J.~E., {et~al.} 2020, \nat,
  586, 228

\bibitem[{{Shannon} {et~al.}(2013){Shannon}, {Blitz}, {Goddard}, \&
  {Heard}}]{Shannon2013}
{Shannon}, R.~J., {Blitz}, M.~A., {Goddard}, A., \& {Heard}, D.~E. 2013, Nature
  Chemistry, 5, 745

\bibitem[{{Sheehan} \& {Eisner}(2017)}]{Sheehan2017}
{Sheehan}, P.~D. \& {Eisner}, J.~A. 2017, \apj, 851, 45

\bibitem[{{Skouteris} {et~al.}(2017){Skouteris}, {Vazart}, {Ceccarelli},
  {Balucani}, {Puzzarini}, \& {Barone}}]{Skouteris2017}
{Skouteris}, D., {Vazart}, F., {Ceccarelli}, C., {et~al.} 2017, \mnras, 468, L1

\bibitem[{{Taquet} {et~al.}(2019){Taquet}, {Bianchi}, {Codella}, {Persson},
  {Ceccarelli}, {Cabrit}, {J{\o}rgensen}, {Kahane}, {L{\'o}pez-Sepulcre}, \&
  {Neri}}]{Taquet2019}
{Taquet}, V., {Bianchi}, E., {Codella}, C., {et~al.} 2019, \aap, 632, A19

\bibitem[{{Taquet} {et~al.}(2012){Taquet}, {Ceccarelli}, \&
  {Kahane}}]{Taquet2012b}
{Taquet}, V., {Ceccarelli}, C., \& {Kahane}, C. 2012, \apjl, 748, L3

\bibitem[{{Taquet} {et~al.}(2015){Taquet}, {L{\'o}pez-Sepulcre}, {Ceccarelli},
  {Neri}, {Kahane}, \& {Charnley}}]{Taquet2015}
{Taquet}, V., {L{\'o}pez-Sepulcre}, A., {Ceccarelli}, C., {et~al.} 2015, \apj,
  804, 81

\bibitem[{{Tobin} {et~al.}(2016){Tobin}, {Looney}, {Li}, {Chandler}, {Dunham},
  {Segura-Cox}, {Sadavoy}, {Melis}, {Harris}, {Kratter}, \&
  {Perez}}]{Tobin2016}
{Tobin}, J.~J., {Looney}, L.~W., {Li}, Z.-Y., {et~al.} 2016, \apj, 818, 73

\bibitem[{{Vigren} {et~al.}(2008){Vigren}, {Kami{\'n}ska}, {Hamberg},
  {Zhaunerchyk}, {Thomas}, {Danielsson}, {Semaniak}, {Andersson}, {Larsson}, \&
  {Geppert}}]{Vigren2008}
{Vigren}, E., {Kami{\'n}ska}, M., {Hamberg}, M., {et~al.} 2008, Physical
  Chemistry Chemical Physics (Incorporating Faraday Transactions), 10, 4014

\bibitem[{{Walmsley} {et~al.}(1989){Walmsley}, {Henkel}, {Jacq}, \&
  {Baudry}}]{Walmsley1989}
{Walmsley}, C.~M., {Henkel}, C., {Jacq}, T., \& {Baudry}, A. 1989, The Physics
  and Chemistry of Interstellar Molecular Clouds - mm and Sub-mm Observations
  in Astrophysics, ed. G.~{Winnewisser} \& J.~T. {Armstrong}, Vol. 331, 107

\bibitem[{{Watanabe} \& {Kouchi}(2002)}]{Watanabe2002}
{Watanabe}, N. \& {Kouchi}, A. 2002, \apjl, 571, L173

\bibitem[{{Yang} {et~al.}(2021){Yang}, {Sakai}, {Zhang}, {Murillo}, {Zhang},
  {Higuchi}, {Zeng}, {L{\'o}pez-Sepulcre}, {Yamamoto}, {Lefloch}, {Bouvier},
  {Ceccarelli}, {Hirota}, {Imai}, {Oya}, {Sakai}, \& {Watanabe}}]{Yang2021}
{Yang}, Y.-L., {Sakai}, N., {Zhang}, Y., {et~al.} 2021, \apj, 910, 20

\bibitem[{{Zucker} {et~al.}(2018){Zucker}, {Schlafly}, {Speagle}, {Green},
  {Portillo}, {Finkbeiner}, \& {Goodman}}]{Zucker2018}
{Zucker}, C., {Schlafly}, E.~F., {Speagle}, J.~S., {et~al.} 2018, \apj, 869, 83

\end{thebibliography}

\clearpage
\onecolumn

\appendix
\section{List of transitions and line properties of the CH$_{\rm 2}$DCN and CH$_{\rm 3}$CN emission.}
\begin{longtable}{lccccccccc}
\caption{List of transitions and line properties (in $T_{\rm MB}$ scale) of the CH$_{\rm 2}$DCN and CH$_{\rm 3}$CN emission.
The columns give the transition and their frequency (GHz),
    the telescope HPBW ($\arcsec$),
    the upper level energy $E_{\rm up}$ (K),
    the S$\mu^{2}$ product (D$^{2}$), 
    the line rms (mK) and its peak temperature (mK), 
    the peak velocities (km/s), 
    the line full width at half maximum (FWHM) (km/s),
and the velocity integrated line intensity $I_{\rm int}$ (mK km/s).}
\\
\hline
\multicolumn{1}{c}{Transition} &
\multicolumn{1}{c}{$\nu$$^{\rm a}$} &
\multicolumn{1}{c}{$HPBW$} &
\multicolumn{1}{c}{$E_{\rm up}$$^{\rm a}$} &
\multicolumn{1}{c}{$S\mu^2$$^{\rm a}$} &
\multicolumn{1}{c}{rms} &
\multicolumn{1}{c}{$T_{\rm peak}$$^{\rm b}$} &
\multicolumn{1}{c}{$V_{\rm peak}$$^{\rm b}$} &
\multicolumn{1}{c}{$FWHM$$^{\rm b}$} &
\multicolumn{1}{c}{$I_{\rm int}$$^{\rm b}$} \\
\multicolumn{1}{c}{ } &
\multicolumn{1}{c}{(GHz)} &
\multicolumn{1}{c}{($\arcsec$)} &
\multicolumn{1}{c}{(K)} &
\multicolumn{1}{c}{(D$^2$)} & 
\multicolumn{1}{c}{(mK)} &
\multicolumn{1}{c}{(mK)} &
\multicolumn{1}{c}{(km s$^{-1}$)} &
\multicolumn{1}{c}{(km s$^{-1}$)} &
\multicolumn{1}{c}{(mK km s$^{-1}$)} \\ 
\hline
\endfirsthead
\hline
\multicolumn{1}{c}{Transition} &
\multicolumn{1}{c}{$\nu$$^{\rm a}$} &
\multicolumn{1}{c}{$HPBW$} &
\multicolumn{1}{c}{$E_{\rm up}$$^{\rm a}$} &
\multicolumn{1}{c}{$S\mu^2$$^{\rm a}$} &
\multicolumn{1}{c}{rms} &
\multicolumn{1}{c}{$T_{\rm peak}$$^{\rm b}$} &
\multicolumn{1}{c}{$V_{\rm peak}$$^{\rm b}$} &
\multicolumn{1}{c}{$FWHM$$^{\rm b}$} &
\multicolumn{1}{c}{$I_{\rm int}$$^{\rm b}$} \\
\multicolumn{1}{c}{ } &
\multicolumn{1}{c}{(GHz)} &
\multicolumn{1}{c}{($\arcsec$)} &
\multicolumn{1}{c}{(K)} &
\multicolumn{1}{c}{(D$^2$)} & 
\multicolumn{1}{c}{(mK)} &
\multicolumn{1}{c}{(mK)} &
\multicolumn{1}{c}{(km s$^{-1}$)} &
\multicolumn{1}{c}{(km s$^{-1}$)} &
\multicolumn{1}{c}{(mK km s$^{-1}$)} \\ 
\hline
\endhead
\multicolumn{10}{c}{CH$_{\rm 3}$CN - ASAI}\\
\hline


CH$_{\rm 3}$CN 5$_{\rm 3}$--4$_{\rm 3}$$^{\rm c}$ & 91.9711 & 27 & 78 & 136 & 3 & 15 (2) & +9.50 (0.33)  & 6.0 (0.8) & 95 (11)\\

CH$_{\rm 3}$CN 5$_{\rm 2}$--4$_{\rm 2}$ & 91.9800 & 27 & 42 & 89 & 2 & 20 (2) & +7.77 (0.09) & 4.7 (0.2) & 100 (4)\\

CH$_{\rm 3}$CN 5$_{\rm 1}$--4$_{\rm 1}$ &   91.9853 & 27 & 20 & 102 & 2 & 26 (2) & +8.23 (0.11) & 4.4 (0.3) & 119 (6)\\

CH$_{\rm 3}$CN 5$_{\rm 0}$--4$_{\rm 0}$ & 91.9871 & 27 & 13 & 106 & 2 & 34 (2) & +8.48 (0.06) & 2.7 (0.2) & 98 (5)\\

CH$_{\rm 3}$CN 6$_{\rm 3}$--5$_{\rm 3}$ & 110.3644 & 22 & 83 & 191 & 2 & 22 (2) & +8.63 (0.09) & 2.8 (0.2) & 65 (4)\\

CH$_{\rm 3}$CN 6$_{\rm 2}$--5$_{\rm 2}$ & 110.3750 & 22 & 47 & 113 & 3 & 29 (3) & +8.73 (0.04) & 3.8 (1.1) & 117 (25)\\

CH$_{\rm 3}$CN 6$_{\rm 1}$--5$_{\rm 1}$ & 110.3814 & 22 & 26 & 124 & 4 & 40 (4) & +8.29 (0.12) & 3.1 (0.4) & 131 (13)\\

CH$_{\rm 3}$CN 6$_{\rm 0}$--5$_{\rm 0}$ & 110.3835 & 22 & 18 & 127 & 4 & 30 (4) & +8.41 (0.16) & 4.1 (0.7) & 131 (17)\\

CH$_{\rm 3}$CN 8$_{\rm 4}$--7$_{\rm 4}$$^{\rm d}$ & 147.1292 & 17 & 146 & 127 & 10 & $\leq$ 24 & - & - & -\\

CH$_{\rm 3}$CN 8$_{\rm 3}$--7$_{\rm 3}$ & 147.1491 & 17 & 96 & 291 & 10 & 43 (14) & +7.86 (0.38) & 5.0 (1.0) & 231 (37)\\

CH$_{\rm 3}$CN 8$_{\rm 2}$--7$_{\rm 2}$ & 147.1632 & 17 & 60 & 159 & 10 & 40 (10) & +8.27 (0.24) & 4.9 (0.6) & 208 (22)\\

CH$_{\rm 3}$CN 8$_{\rm 1}$--7$_{\rm 1}$ & 147.1718 & 17 & 39 & 167 & 10 & 60 (10) & +8.28 (0.14) & 3.5 (0.4) & 223 (19)\\

CH$_{\rm 3}$CN 8$_{\rm 0}$--7$_{\rm 0}$ & 147.1746 & 17 & 32 & 170 & 10 & 67 (10) & +8.61 (0.13) & 3.3 (0.5) & 231 (25)\\

CH$_{\rm 3}$CN 9$_{\rm 4}$--8$_{\rm 4}$$^{\rm d}$& 165.5181 & 15 & 154 & 153 & 10 & $\leq$ 28 & - & - & -\\

CH$_{\rm 3}$CN 9$_{\rm 3}$--8$_{\rm 3}$ & 165.5404  & 15 & 104 & 339 & 10 & 43 (10) & +8.39 (0.20) & 4.8 (0.4) & 222(18)\\

CH$_{\rm 3}$CN 9$_{\rm 2}$--8$_{\rm 2}$ & 165.5563  & 15 & 68 & 181 & 10 & 55(10) & +8.27 (0.17) & 3.5 (0.5) & 209 (24)\\

CH$_{\rm 3}$CN 9$_{\rm 1}$--8$_{\rm 1}$ & 165.5659  & 15 & 47 & 188 & 8 & 77 (8) & +8.33 (0.09) & 3.2 (0.2) & 261 (16)\\

CH$_{\rm 3}$CN 9$_{\rm 0}$--8$_{\rm 0}$ & 165.5691   & 15 & 40 & 191 & 8 & 52 (8) & +8.34 (0.14) & 4.8 (0.4) & 264 (19)\\


CH$_{\rm 3}$CN 11$_{\rm 6}$--10$_{\rm 6}$$^{\rm d}$& 202.2154 & 12 & 315 & 328 & 17 & $\leq$ 36 & - & - & -\\

CH$_{\rm 3}$CN 11$_{\rm 5}$--10$_{\rm 5}$$^{\rm d}$& 202.2582 & 12 & 237 & 185 & 16 & $\leq$ 45 & - & - & -\\

CH$_{\rm 3}$CN 11$_{\rm 4}$--10$_{\rm 4}$ & 202.2932   & 12 & 173 & 202 & 16 & 60 (14) & +8.18 (0.19) & 3.7 (0.4) & 236 (24)\\

CH$_{\rm 3}$CN 11$_{\rm 3}$--10$_{\rm 3}$ & 202.3204    & 12 & 123 & 432 & 18 & 117 (20) & +8.21 (0.11) &3.9 (0.2) & 478 (28)\\

CH$_{\rm 3}$CN 11$_{\rm 2}$--10$_{\rm 2}$ & 202.3399   & 12 & 87 & 225 & 16 & 98 (16) & +8.16 (0.30) & 4.6 (0.8) & 475 (65)\\

CH$_{\rm 3}$CN 11$_{\rm 1}$--10$_{\rm 1}$ & 202.3516  & 12 & 65 & 231 & 14 & 101 (17) & +8.08 (0.10) & 3.7 (0.2) & 399 (22)\\

CH$_{\rm 3}$CN 11$_{\rm 0}$--10$_{\rm 0}$ & 202.3555  & 12 & 58 & 233 & 18 & 120 (19) & +8.30 (0.10) & 3.0 (0.3) & 383 (27)\\

CH$_{\rm 3}$CN 12$_{\rm 7}$--11$_{\rm 7}$$^{\rm d}$& 220.5393 & 11 & 419 & 168 & 8 & $\leq$ 23 & - & - & -\\

CH$_{\rm 3}$CN 12$_{\rm 6}$--11$_{\rm 6}$ & 220.5944  & 11 & 326 & 382 & 7 & 53 (7) & +8.58 (0.09) & 4.3 (0.3) & 245 (12) \\

CH$_{\rm 3}$CN 12$_{\rm 5}$--11$_{\rm 5}$ & 220.6411  & 11 & 247 & 210 & 7 & 49 (7) & +8.27 (0.11) & 4.4 (0.2) & 226 (11)\\

CH$_{\rm 3}$CN 12$_{\rm 4}$--11$_{\rm 4}$ & 220.6793 & 11 & 183 & 226 & 9 & 75 (9) & +8.00 (0.11) & 5.5 (0.3) & 439 (19)\\

CH$_{\rm 3}$CN 12$_{\rm 3}$--11$_{\rm 3}$ & 220.7090 & 11 & 133 & 477 & 8 & 99 (8) & +8.26 (0.07) & 5.0 (0.2) & 525 (16)\\

CH$_{\rm 3}$CN 12$_{\rm 2}$--11$_{\rm 2}$ & 220.7303 & 11 & 97 & 247 & 7 & 90 (7) & +8.19 (0.06) & 4.3 (0.1) & 407 (12)\\

CH$_{\rm 3}$CN 12$_{\rm 1}$--11$_{\rm 1}$ & 220.7430  & 11 & 76 & 253 & 7 & 103 (7) & +8.26 (0.05) & 3.9 (0.1) & 490 (6)\\

CH$_{\rm 3}$CN 12$_{\rm 0}$--11$_{\rm 0}$ & 220.7473  & 11 &69 & 254 & 7 & 105 (7) & +8.21 (0.02) & 4.4 (0.2) & 488 (18)\\

CH$_{\rm 3}$CN 13$_{\rm 7}$--12$_{\rm 7}$ & 238.9127  & 10 & 430 & 196 & 8 & 37 (8) & +7.90 (0.15) & 4.3 (0.4) & 170 (12)\\

CH$_{\rm 3}$CN 13$_{\rm 6}$--12$_{\rm 6}$ & 238.9724  & 10 & 337 & 434 & 9 & 59 (9) & +7.79 (0.11) & 3.9 (0.2) & 248 (14)\\


CH$_{\rm 3}$CN 13$_{\rm 5}$--12$_{\rm 5}$$^{\rm c}$ & 239.0229 & 10 & 259 & 235 & 9 & 72 (9) & +8.81 (0.13) & 7.3 (0.3) & 560 (21)\\

CH$_{\rm 3}$CN 13$_{\rm 4}$--12$_{\rm 4}$ & 239.0643   & 10 & 195 & 249 & 9 & 72 (9) & +7.80 (0.10) & 4.8 (0.2) & 365 (15)\\

CH$_{\rm 3}$CN 13$_{\rm 3}$--12$_{\rm 3}$ & 239.0965   & 10 & 145 & 522 & 10 & 105 (10) & +7.97 (0.08) & 4.8 (0.2) & 532 (18)\\

CH$_{\rm 3}$CN 13$_{\rm 2}$--12$_{\rm 2}$ & 239.1195  & 10 & 109 & 269 & 11 & 98 (11) & +8.07 (0.08) & 4.3 (0.2) & 445 (17)\\


CH$_{\rm 3}$CN 13$_{\rm 1}$--12$_{\rm 1}$$^{\rm e}$ & 239.1333 & 10 & 87 & 274 & 9 & - & - & - & -\\
CH$_{\rm 3}$CN 13$_{\rm 0}$--12$_{\rm 0}$$^{\rm e}$ & 239.1379 & 10 & 80 & 276 & 9 & - & - & - & -\\

CH$_{\rm 3}$CN 14$_{\rm 7}$--13$_{\rm 7}$ & 257.2849  & 10 & 442 & 223 & 6 & 35 (6) & +8.08 (0.12) & 3.9 (0.3) & 146 (9) \\

CH$_{\rm 3}$CN 14$_{\rm 6}$--13$_{\rm 6}$ & 257.3492 & 10 & 350 & 484 & 6 & 55 (6) & +8.58 (0.09) & 4.9 (0.2) & 290 (11) \\

CH$_{\rm 3}$CN 14$_{\rm 5}$--13$_{\rm 5}$$^{\rm f}$ & 257.4036 & 10 & 271 & 259 & 6 & - & - & - &-\\


CH$_{\rm 3}$CN 14$_{\rm 4}$--13$_{\rm 4}$ & 257.4481 & 10 & 207 & 273 & 7 & 76 (7) & +8.33 (0.06) & 4.4 (0.2) & 360 (10)\\

CH$_{\rm 3}$CN 14$_{\rm 3}$--13$_{\rm 3}$ & 257.4828 & 10 & 157 & 566 & 5 & 97 (5) & +8.38 (0.04) & 4.5 (0.1) & 462 (9)\\

CH$_{\rm 3}$CN 14$_{\rm 2}$--13$_{\rm 2}$ & 257.5076 & 10 & 121 & 291 & 6 & 86 (6) & +8.65 (0.05) & 4.2 (0.1) & 385 (10)\\

CH$_{\rm 3}$CN 14$_{\rm 1}$--13$_{\rm 1}$ & 257.5224 & 10 & 100 & 295 & 8 & 93 (8) & +8.44 (0.09) & 4.7 (0.2) & 467 (15)\\

CH$_{\rm 3}$CN 14$_{\rm 0}$--13$_{\rm 0}$ & 257.5274 & 10 & 93 & 297 & 8 & 103 (8) & +8.55 (0.52) & 4.3 (0.4) & 468 (15)\\





\hline
\multicolumn{10}{c}{CH$_{\rm 2}$DCN - SOLIS \& ASAI}\\
\hline

CH$_{\rm 2}$DCN 5$_{\rm 1,4}$--4$_{\rm 1,3}$$^{\rm g}$ & 87.2115 & 1.5 & 18 & 74 & 38 & 306 (25) & +9 (7) & -- & 2226  (140)\\ 

CH$_{\rm 2}$DCN 6$_{\rm 1,6}$--5$_{\rm 1,5}$$^{\rm g}$ & 103.7486 & 1.3 & 23 & 90 & 26 & 290 (30) & +9 (6) & -- & 1836 (200)\\ 

\hline
\vspace{0.15cm}
CH$_{\rm 2}$DCN 12$_{\rm 1,12}$--11$_{\rm 1,11}$$^{\rm c}$ & 207.4715 & 12 & 70 & 183 & 8 & - & - & - & - \\


CH$_{\rm 2}$DCN 12$_{\rm 6,6}$--11$_{\rm 6,5}$ $^{\rm h}$& \multirow{2}{*}{208.2668} &  \multirow{2}{*}{12} &  \multirow{2}{*}{259} & \multirow{2}{*}{138} & \multirow{2}{*}{8} &\multirow{2}{*}{-} & \multirow{2}{*}{-} & \multirow{2}{*}{-} & \multirow{2}{*}{-}\\
\vspace{0.15cm}
CH$_{\rm 2}$DCN 12$_{\rm 6,7}$--11$_{\rm 6,6}$$^{\rm h}$ \\

CH$_{\rm 2}$DCN 12$_{\rm 5,7}$--11$_{\rm 5,6}$$^{\rm i}$ & \multirow{2}{*}{208.3052} &  \multirow{2}{*}{12} &  \multirow{2}{*}{200} &  \multirow{2}{*}{152} & \multirow{2}{*}{7} & \multirow{2}{*}{23 (4)} & \multirow{2}{*}{+8.34 (0.36)} & \multirow{2}{*}{1.6 (0.8)} & \multirow{2}{*}{38 (17)}\\
\vspace{0.15cm}
CH$_{\rm 2}$DCN 12$_{\rm 5,8}$--11$_{\rm 5,7}$$^{\rm i}$ \\

CH$_{\rm 2}$DCN 12$_{\rm 4,9}$--11$_{\rm 4,8}$$^{\rm d}$ & \multirow{2}{*}{208.3372} &  \multirow{2}{*}{12} & \multirow{2}{*}{151} & \multirow{2}{*}{164} & \multirow{2}{*}{8} & \multirow{2}{*}{$\leq$ 19} & \multirow{2}{*}{-} & \multirow{2}{*}{-} & \multirow{2}{*}{-}\\
\vspace{0.15cm}
CH$_{\rm 2}$DCN 12$_{\rm 4,8}$--11$_{\rm 4,7}$$^{\rm d}$ \\

\vspace{0.15cm}
CH$_{\rm 2}$DCN 12$_{\rm 0,12}$--11$_{\rm 0,11}$$^{\rm c}$ & 208.3456 & 12 & 65 & 184 & 8 & - & - & - & - \\

CH$_{\rm 2}$DCN 12$_{\rm 3,10}$--11$_{\rm 3,9}$$^{\rm d}$ & \multirow{2}{*}{208.3636} &  \multirow{2}{*}{12} &  \multirow{2}{*}{114} & \multirow{2}{*}{173} & \multirow{2}{*}{8} & \multirow{2}{*}{$\leq$ 23} & \multirow{2}{*}{-} & \multirow{2}{*}{-} & \multirow{2}{*}{-}\\
\vspace{0.15cm}
CH$_{\rm 2}$DCN 12$_{\rm 3,9}$--11$_{\rm 3,8}$$^{\rm d}$ \\

CH$_{\rm 2}$DCN 12$_{\rm 2,11}$--11$_{\rm 2,10}$$^{\rm d}$ & 208.3684 & 12 & 87 & 179 & 8 & $\leq$ 23 & - & - & - \\

CH$_{\rm 2}$DCN 12$_{\rm 2,10}$--11$_{\rm 2,9}$$^{\rm d}$ & 208.4116 & 12 & 87 & 179 & 8 & $\leq$ 21 & - & - & - \\

CH$_{\rm 2}$DCN 12$_{\rm 1,11}$--11$_{\rm 1,10}$ &  209.2781 & 12 & 70 & 183 & 7 & 26 (8) & +8.65 (0.32) & 4.3 (0.7) & 118 (18) \\  

\vspace{0.15cm}

CH$_{\rm 2}$DCN 13$_{\rm 1,13}$--12$_{\rm 1,12}$$^{\rm d}$ & 224.7545 & 11 & 81 & 199 & 8 & $\leq$ 23 & - & - & - \\

CH$_{\rm 2}$DCN 13$_{\rm 6,7}$--12$_{\rm 6,6}$$^{\rm j}$& \multirow{2}{*}{225.6182} &  \multirow{2}{*}{11} & \multirow{2}{*}{270} & \multirow{2}{*}{157} & \multirow{2}{*}{8} & \multirow{2}{*}{-} & \multirow{2}{*}{-} & \multirow{2}{*}{-} & \multirow{2}{*}{-}\\
\vspace{0.15cm}

CH$_{\rm 2}$DCN 13$_{\rm 6,8}$--12$_{\rm 6,7}$$^{\rm j}$ \\

CH$_{\rm 2}$DCN 13$_{\rm 5,8}$--12$_{\rm 5,7}$$^{\rm d}$& \multirow{2}{*}{225.6598} &  \multirow{2}{*}{11} &  \multirow{2}{*}{211} &  \multirow{2}{*}{170} & \multirow{2}{*}{8} & \multirow{2}{*}{$\leq$ 18} & \multirow{2}{*}{-} & \multirow{2}{*}{-} & \multirow{2}{*}{-}\\
\vspace{0.15cm}

CH$_{\rm 2}$DCN 13$_{\rm 5,9}$--12$_{\rm 5,8}$$^{\rm d}$ \\

CH$_{\rm 2}$DCN 13$_{\rm 4,10}$--12$_{\rm 4,9}$$^{\rm k}$& \multirow{2}{*}{225.6946} &  \multirow{2}{*}{11} &  \multirow{2}{*}{162} & \multirow{2}{*}{181} & \multirow{2}{*}{13} & \multirow{2}{*}{-} & \multirow{2}{*}{-} & \multirow{2}{*}{-} & \multirow{2}{*}{-}\\
\vspace{0.15cm}

CH$_{\rm 2}$DCN 13$_{\rm 4,9}$--12$_{\rm 4,8}$$^{\rm k}$ \\
\vspace{0.15cm}

CH$_{\rm 2}$DCN 13$_{\rm 0,13}$--12$_{\rm 0,12}$$^{\rm k}$ & 225.6951 & 11 & 76 & 200 & 13 & - & - & - & - \\

CH$_{\rm 2}$DCN 13$_{\rm 3,11}$--12$_{\rm 3,10}$$^{\rm e}$& 225.7238 & \multirow{2}{*}{11} &  \multirow{2}{*}{124} & \multirow{2}{*}{189} & \multirow{3}{*}{9} & \multirow{3}{*}{-} & \multirow{3}{*}{-} & \multirow{3}{*}{-} & \multirow{3}{*}{-}\\
CH$_{\rm 2}$DCN 13$_{\rm 3,10}$--12$_{\rm 3,9}$$^{\rm e}$ & 225.7241 \\
\vspace{0.15cm}

CH$_{\rm 2}$DCN 13$_{\rm 2,12}$--12$_{\rm 2,11}$$^{\rm e}$ & 225.7265 & 11 & 97 & 195\\

CH$_{\rm 2}$DCN 13$_{\rm 2,11}$--12$_{\rm 2,10}$ & 225.7815 & 11 & 97 & 195 & 7 & 30 (10) & +8.59 (0.13) & 1.8 (1.0) & 60 (23)\\

CH$_{\rm 2}$DCN 13$_{\rm 1,12}$--12$_{\rm 1,11}$$^{\rm l}$ & 226.7113 & 11 & 82 & 199 & 8 & - & - & - & - \\
\vspace{0.15cm}

CH$_{\rm 2}$DCN 14$_{\rm 1,14}$--13$_{\rm 1,13}$$^{\rm m}$ & 242.0358 & 10 & 93 & 214 & 8 & - & - & - & - \\

CH$_{\rm 2}$DCN 14$_{\rm 7,7}$--13$_{\rm 7,6}$$^{\rm n}$& \multirow{2}{*}{242.9159} &  \multirow{2}{*}{10} & \multirow{2}{*}{351} & \multirow{2}{*}{161} & \multirow{2}{*}{11} & \multirow{2}{*}{-} & \multirow{2}{*}{-} & \multirow{2}{*}{-} & \multirow{2}{*}{-}\\
\vspace{0.15cm}

CH$_{\rm 2}$DCN 14$_{\rm 7,8}$--13$_{\rm 7,7}$$^{\rm n}$ \\

CH$_{\rm 2}$DCN 14$_{\rm 6,8}$--13$_{\rm 6,7}$$^{\rm d}$& \multirow{2}{*}{242.9685} &  \multirow{2}{*}{10} & \multirow{2}{*}{281} & \multirow{2}{*}{175} &\multirow{2}{*}{11} & \multirow{2}{*}{$\leq$ 20} & \multirow{2}{*}{-} & \multirow{2}{*}{-} & \multirow{2}{*}{-}\\
\vspace{0.15cm}

CH$_{\rm 2}$DCN 14$_{\rm 6,9}$--13$_{\rm 6,8}$$^{\rm d}$ \\

CH$_{\rm 2}$DCN 14$_{\rm 5,9}$--13$_{\rm 5,8}$$^{\rm d}$& \multirow{2}{*}{243.0134} &  \multirow{2}{*}{10} & \multirow{2}{*}{222} & \multirow{2}{*}{188} & \multirow{2}{*}{10} &\multirow{2}{*}{$\leq$ 21} & \multirow{2}{*}{-} & \multirow{2}{*}{-} & \multirow{2}{*}{-}\\
\vspace{0.15cm}

CH$_{\rm 2}$DCN 14$_{\rm 5,10}$--13$_{\rm 5,9}$$^{\rm d}$\\

\vspace{0.15cm}

CH$_{\rm 2}$DCN 14$_{\rm 0,14}$--13$_{\rm 0,13}$ & 243.0415 & 10 & 87 & 215 & 8 & 31 (8) & +8.84 (0.27) & 4.3 (1.1) & 140 (30)\\
CH$_{\rm 2}$DCN 14$_{\rm 4,11}$--13$_{\rm 4,10}$ & \multirow{2}{*}{243.0512} &  \multirow{2}{*}{10} &  \multirow{2}{*}{174} &  \multirow{2}{*}{198} & \multirow{2}{*}{10} & \multirow{2}{*}{32 (6)} & \multirow{2}{*}{+8.79 (0.39)} & \multirow{2}{*}{4.0 (1.0)} & \multirow{2}{*}{134 (27)}\\
\vspace{0.15cm}

CH$_{\rm 2}$DCN 14$_{\rm 4,10}$--13$_{\rm 4,9}$ \\

CH$_{\rm 2}$DCN 14$_{\rm 3,12}$--13$_{\rm 3,11}$ & 243.0830 & 10 & 136 & 205 & 8 &\multirow{3}{*}{40 (5)} &\multirow{3}{*}{+8.42 (0.21)} &\multirow{3}{*}{4.6 (0.5)} &\multirow{3}{*}{197 (17)}\\
CH$_{\rm 2}$DCN 14$_{\rm 2,13}$--13$_{\rm 2,12}$ & 243.0833 & 10 & 109 & 211  & 8  \\
\vspace{0.15cm}

CH$_{\rm 2}$DCN 14$_{\rm 3,11}$--13$_{\rm 3,10}$ & 243.0835 & 10 & 136 & 205  & 8\\

CH$_{\rm 2}$DCN 14$_{\rm 2,12}$--13$_{\rm 2,11}$$^{\rm d}$ & 243.1520 & 10 & 109 & 211 & 8 & $\leq$ 20 & - & - & - \\

CH$_{\rm 2}$DCN 14$_{\rm 1,13}$--13$_{\rm 1,12}$ & 244.1428 &  10 & 93 & 214 & 9 & 36 (7) & +9.26 (0.13) & 2.4 (0.5) & 90 (14) \\

\hline
\label{table:CH3CN-lines}
\end{longtable}
\vspace{-0.3cm}
\begin{minipage}{17.5cm}
\footnotesize{$^{\rm a}$ Frequencies and spectroscopic parameters have been provided by \citet{Cazzoli2006} and \citet{Muller2009} and by the Jet Propulsion Laboratory molecular database \citet{Pickett1998} for CH$_3$CN, while they were provided by \citet{Muller2009} and retrieved from the Cologne Database for Molecular Spectroscopy \citep{Muller2005} for CH$_{\rm 2}$DCN.
$^{\rm b}$ The errors in brackets are the Gaussian fit uncertainties.
$^{\rm c}$ Blended with unidentified line (see Fig. \ref{Fig:spectra+model1} and Fig. \ref{Fig:spectra+model2}) and thus excluded from the further analysis.
$^{\rm d}$ Excluded from further analysis since the line peak is below the detection threshold (S/N $>$ 3$\sigma$ see Sec. \ref{subsec:obs-asai}).
$^{\rm e}$ Conservatively excluded from the analysis since the line profiles are broad (composed of different transitions close in frequency, see Fig. \ref{Fig:spectra+model1} and \ref{Fig:spectra+model2}).
$^{\rm f}$ Blended with CH$_{\rm 3}$OH 18$_{\rm 3, 16}$--18$_{\rm 2, 17}$ A, 257.4021 GHz, E$_{\rm up}$= 447 K.
$^{\rm g}$ These transitions are only detected by the NOEMA/SOLIS observations.
The low spectral resolution (6--7 km s$^{-1}$) does not allow us to safely estimate the line width. A Gaussian fit is not performed given the unresolved line profiles. We note that all the expected CH$_{\rm 2}$DCN transitions are detected in the NOEMA/SOLIS dataset, but we consider conservatively only these two lines which are not contaminated by other molecular species.
$^{\rm h}$ Blended with CH$_{\rm 3}$CHO 11$_{\rm 0, 11}$--10$_{\rm 0, 10}$ A, 208.2670 GHz, E$_{\rm up}$= 60 K.
$^{\rm i}$ These transitions have the highest upper level energy of all the detected transitions, while others with similar intensities and lower energies are not detected. To be conservative we decided to exclude the line from the analysis.
$^{\rm j}$Blended with HCOOCH$_{\rm 3}$ 19$_{\rm 3, 17}$--18$_{\rm 3, 16}$ A, 225.6187 GHz, E$_{\rm up}$= 117 K.
$^{\rm k}$Blended with H$_{\rm 2}$CO 3$_{\rm 1, 2}$--2$_{\rm 1, 1}$, 225.6978 GHz, E$_{\rm up}$= 33 K.
$^{\rm l}$ Blended with HCOOCH$_{\rm 3}$ 20$_{\rm 2, 19}$--19$_{\rm 2, 18}$ E, 226.7131 GHz, E$_{\rm up}$= 120 K.
$^{\rm m}$ Blended with CH$_{\rm 2}$DOH 11$_{\rm 2, 9}$--11$_{\rm 1, 10}$ o1, 242.0336 GHz, E$_{\rm up}$= 177 K.
$^{\rm n}$ Blended with C$^{\rm 33}$S 5--4, 242.9136 GHz, E$_{\rm up}$= 35 K.}
\end{minipage}
\section{ASAI versus SOLIS spectra of CH$_2$DCN}

\begin{figure}[h]
\begin{center}
\includegraphics[angle=0,width=8cm]{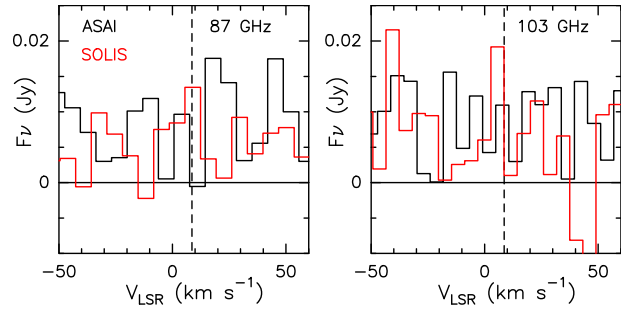} 
  \caption{Comparison of the CH$_{\rm 2}$DCN emission in the ASAI and SOLIS observations. The data have been resampled to match the same spectral resolution. }
 \label{Fig:asai-solis}
  \end{center}
\end{figure}

Figure \ref{Fig:asai-solis} shows the comparison between the IRAM-30m 3mm spectrum obtained in the context of the ASAI Large Program \citet{Lefloch2018} and the spectra derived by integrating the emission
in the NOEMA-SOLIS images in a region equal to the HPBW of the IRAM-30m
(28$\arcsec$ at 87 GHz, and 24$\arcsec$ at 103 GHz).

\section{ASAI spectra}

Figures \ref{Fig:spectra+model1} and  \ref{Fig:spectra+model2} show the ASAI spectra overlaid with the synthetic spectra derived for the CH$_{\rm 3}$CN  and CH$_{\rm 2}$DCN  emission. The  best fit model (see Sects. \ref{subsec:nonLTE} and \ref{subsec:CH2DCN-LTE}) and  the uncertainties are shown.

\begin{figure*}[h]
\begin{center}
\includegraphics[angle=0,width=15cm]{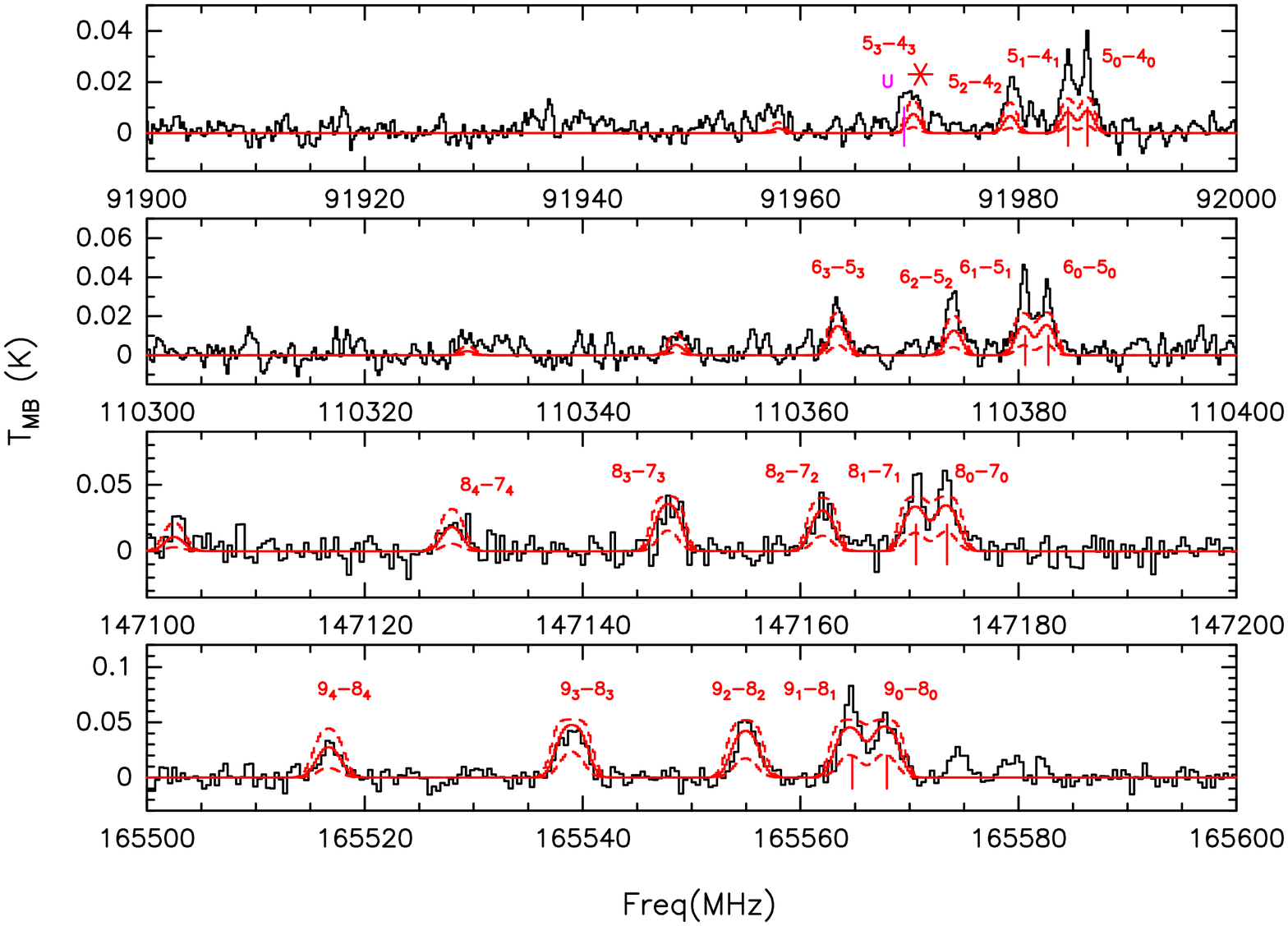}
\includegraphics[angle=0,width=15cm]{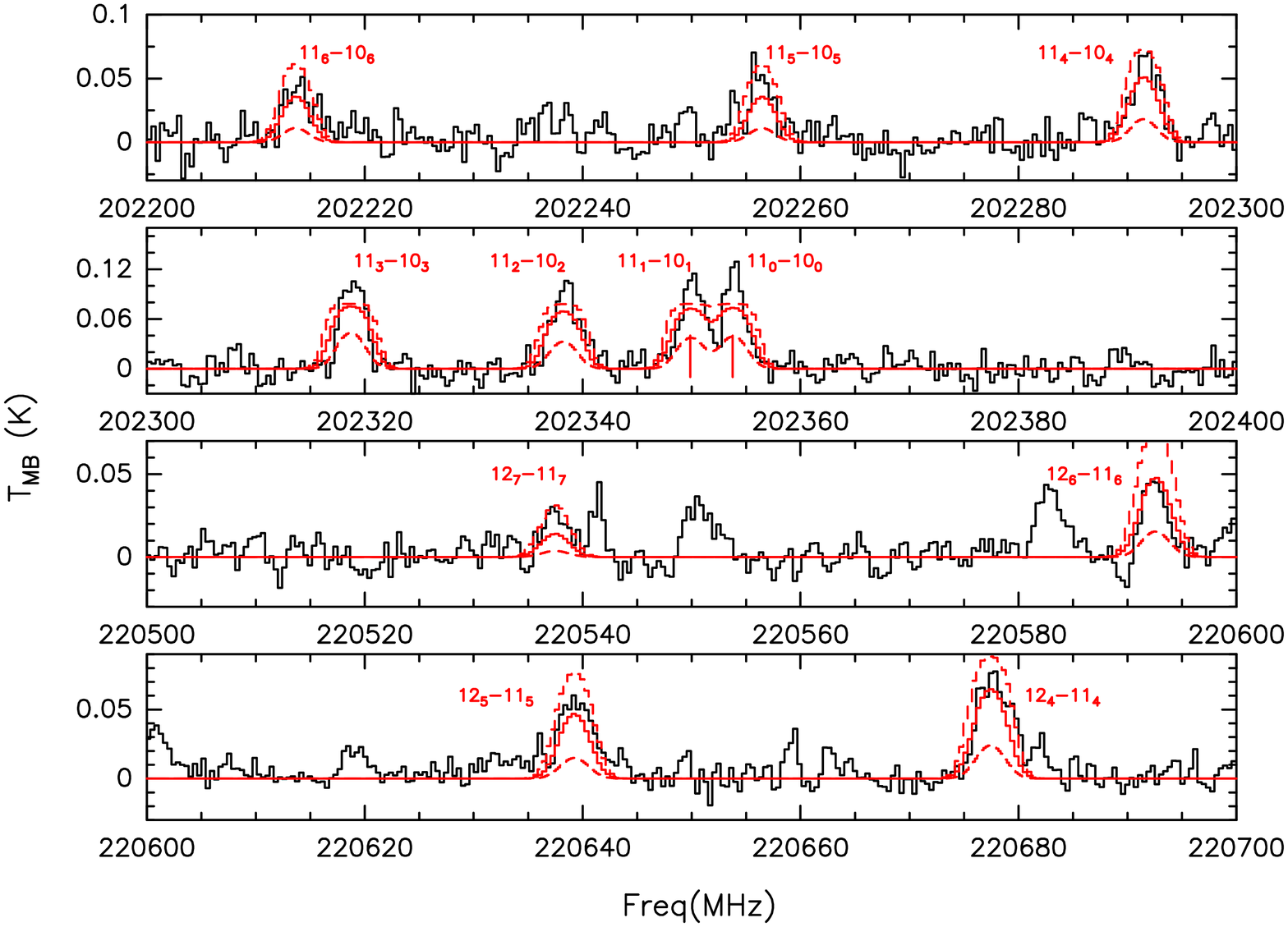} 
  \caption{Synthetic CH$_{\rm 3}$CN spectra (in red)  overlaid to the ASAI dataset at 1.3mm. The continuous lines show the best fit model (see Sect. \ref{subsec:nonLTE}), while the dashed lines take into account the uncertainties. The asterisk denotes the line contaminated CH$_{\rm 3}$CN profiles. Synthetic spectra are plotted using the CLASS Weeds package \citep{Maret2011}. Spectra are smoothed to a spectral resolution of 0.5 km s$^{-1}$.}
 \label{Fig:spectra+model1}
  \end{center}
\end{figure*}

\begin{figure*}[h]
\addtocounter{figure}{-1}
\begin{center}
\includegraphics[angle=0,width=15cm]{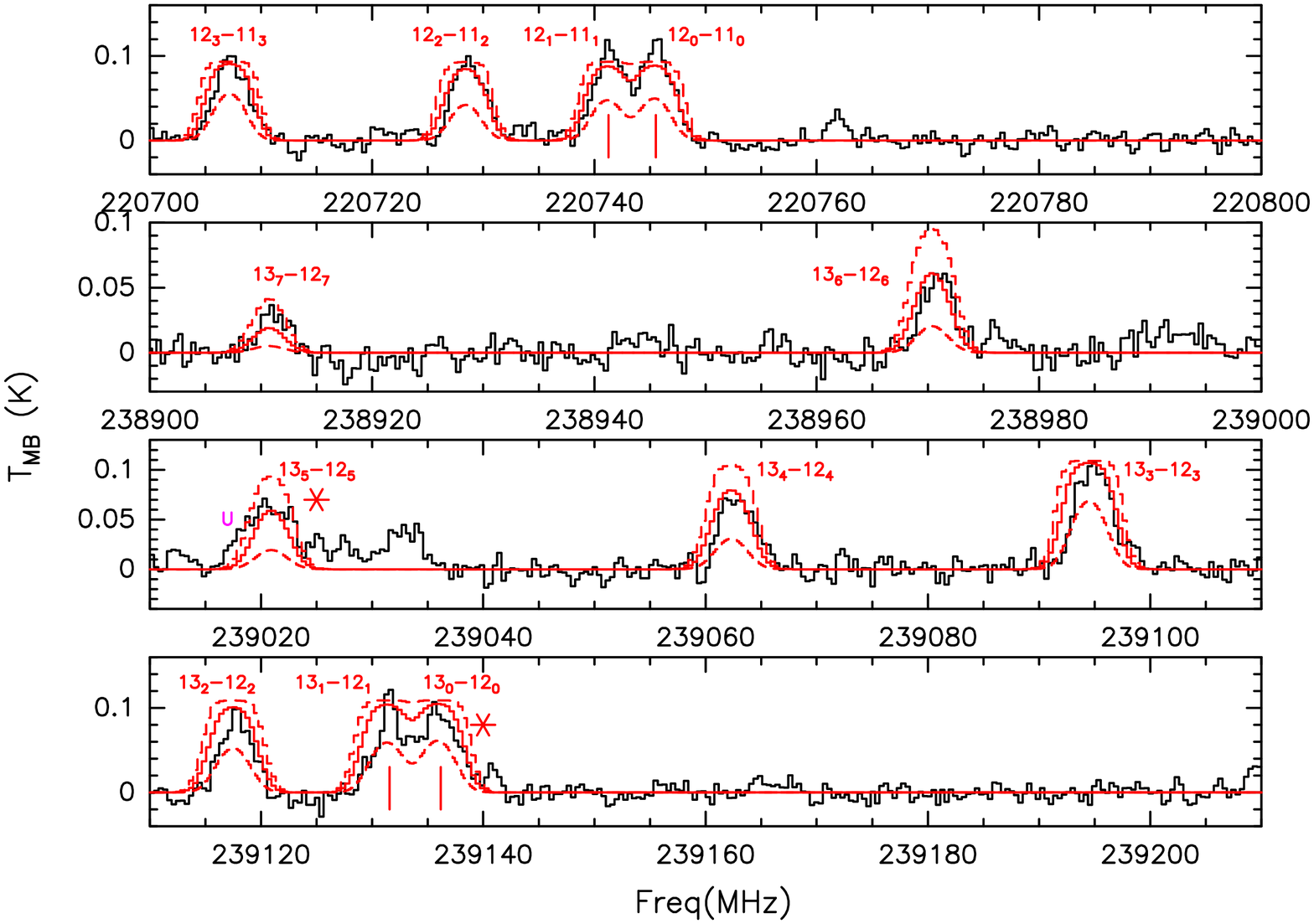}
\includegraphics[angle=0,width=15cm]{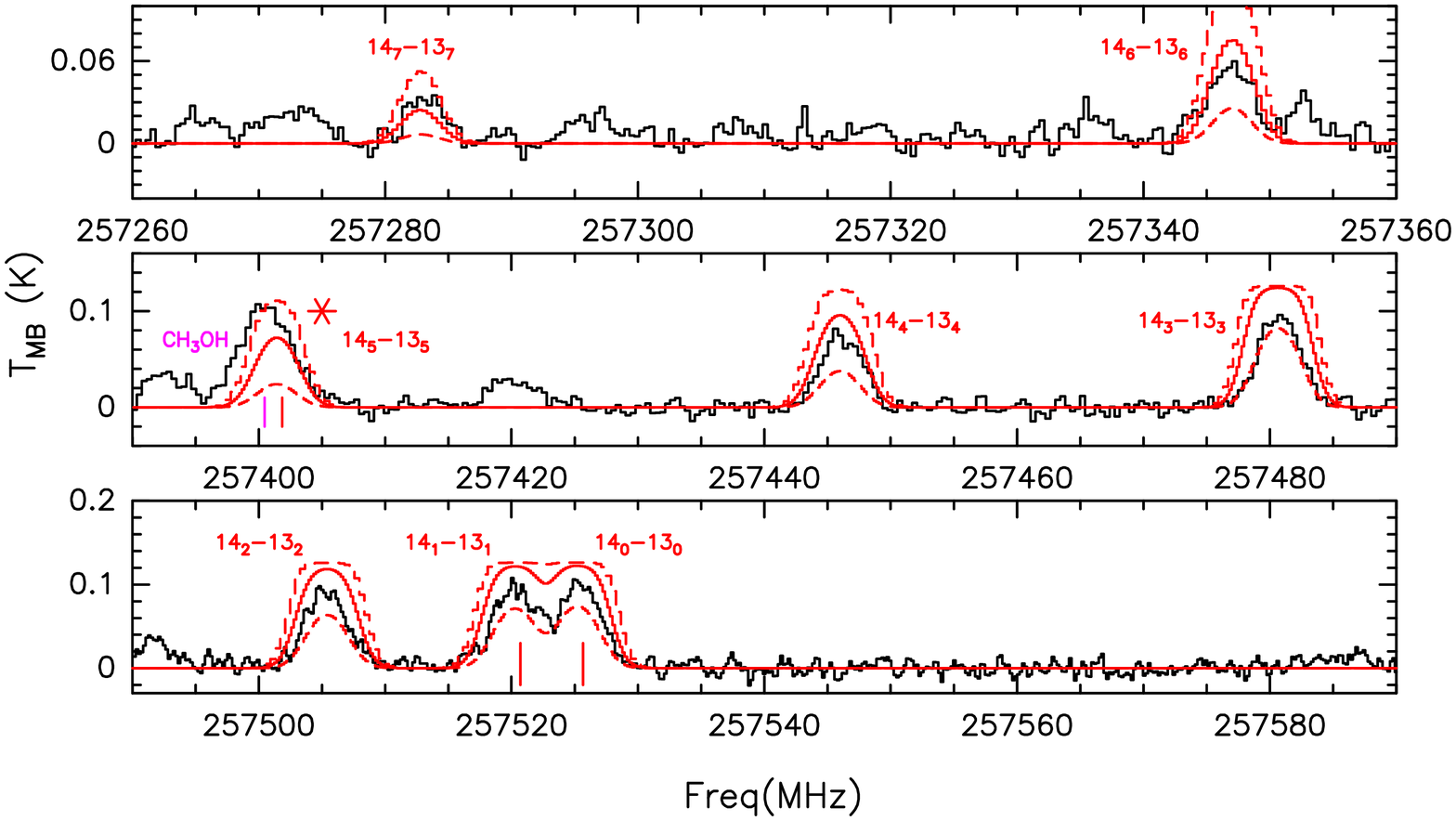} 
  \caption{\textit{Continued.}}
  \end{center}
\end{figure*}

\begin{figure*}[h]
\begin{center}
\includegraphics[angle=0,width=15cm]{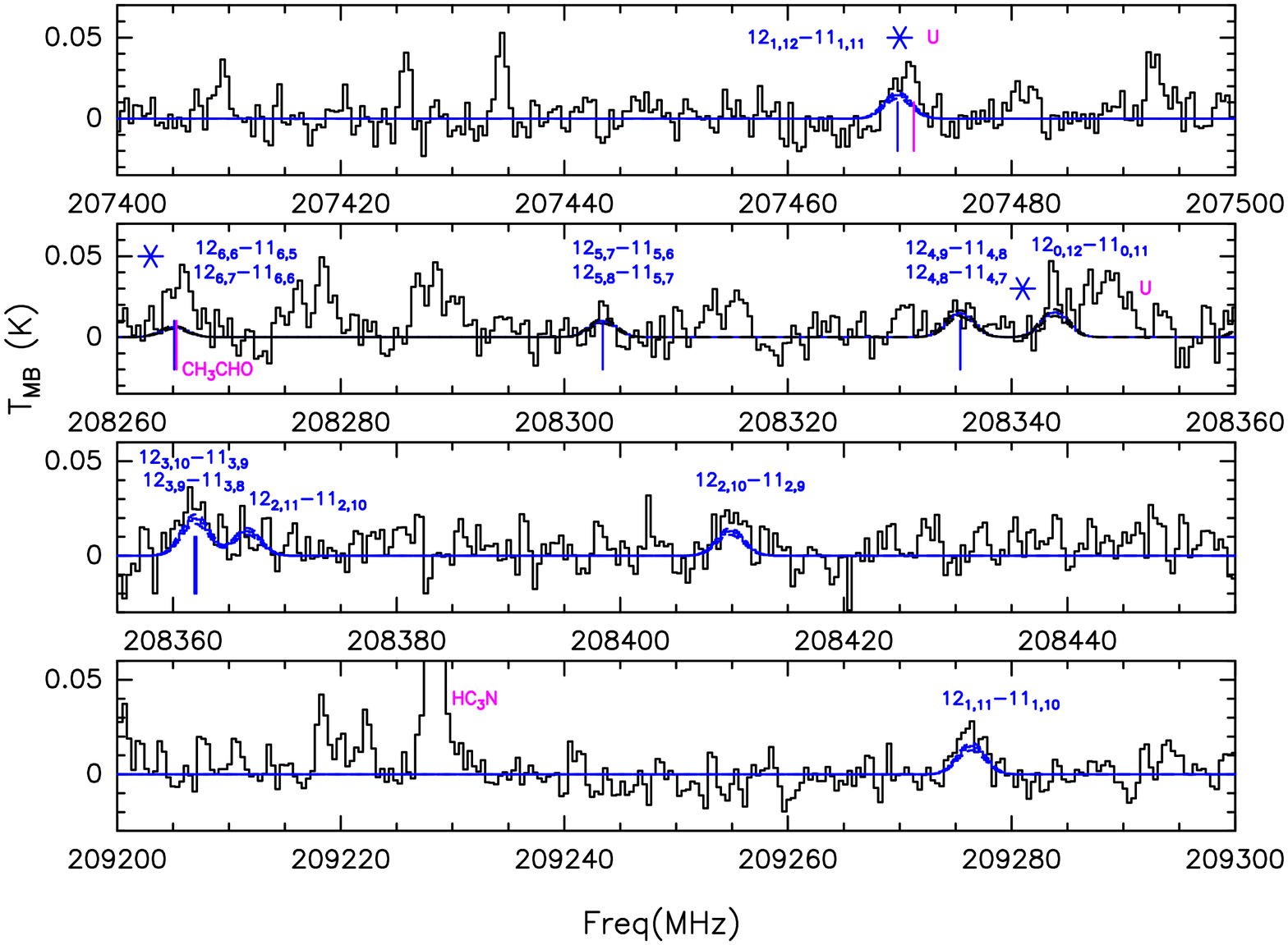}
\includegraphics[angle=0,width=15cm]{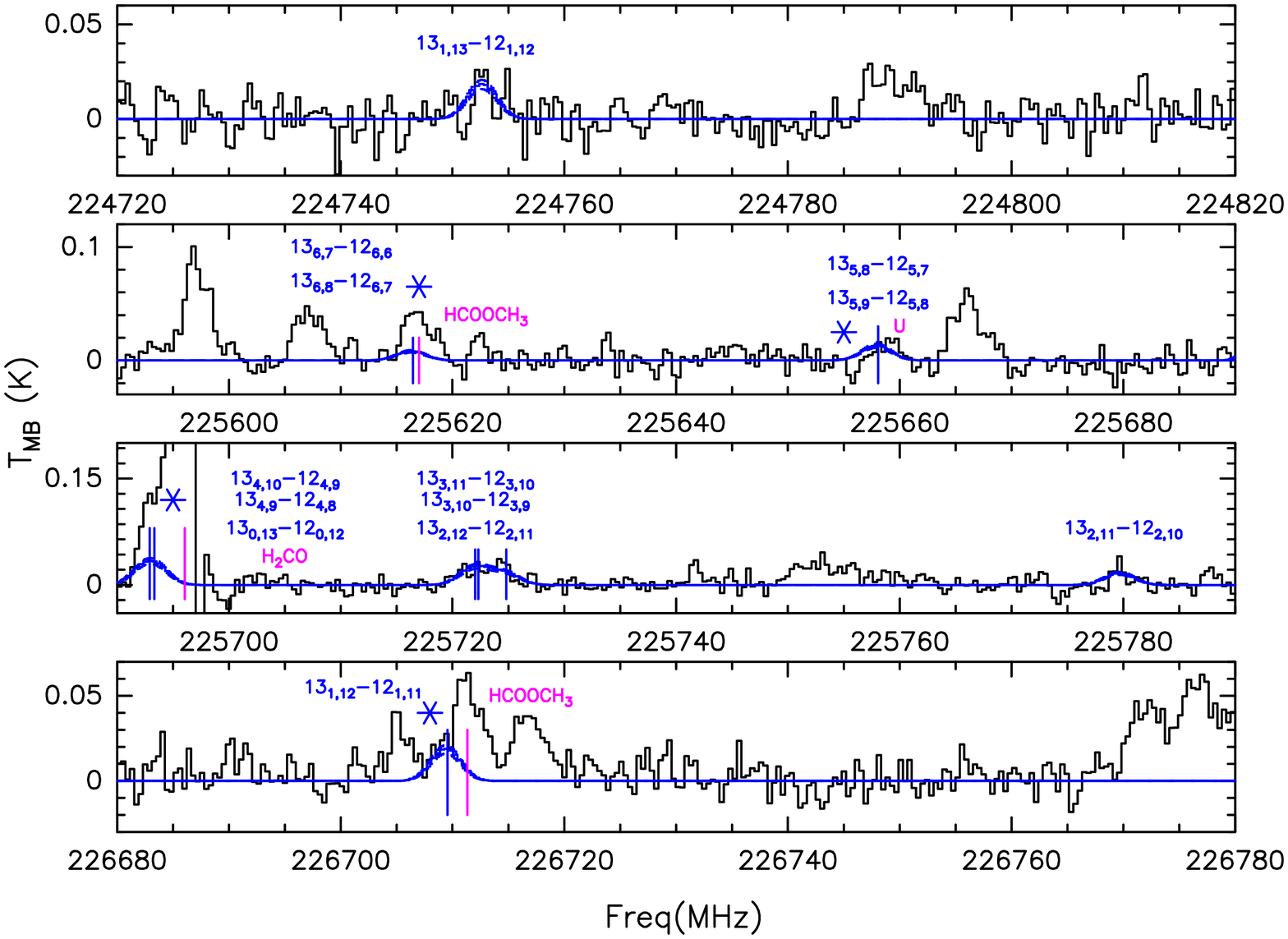} 
  \caption{Synthetic CH$_{\rm 2}$DCN (in blue) spectra overlaid to the ASAI dataset at 1.3mm. The continuous lines show the best fit model (see Sect. \ref{subsec:CH2DCN-LTE}), while the dashed lines take into account the uncertainties. The asterisk denotes the line contaminated CH$_{\rm 2}$DCN profiles. Synthetic spectra are plotted using the CLASS Weeds package \citep{Maret2011}. Spectra are smoothed to a spectral resolution of 0.5 km s$^{-1}$.}
 \label{Fig:spectra+model2}
  \end{center}
\end{figure*}

\begin{figure*}[h]
\addtocounter{figure}{-1}
\begin{center}
\includegraphics[angle=0,width=15cm]{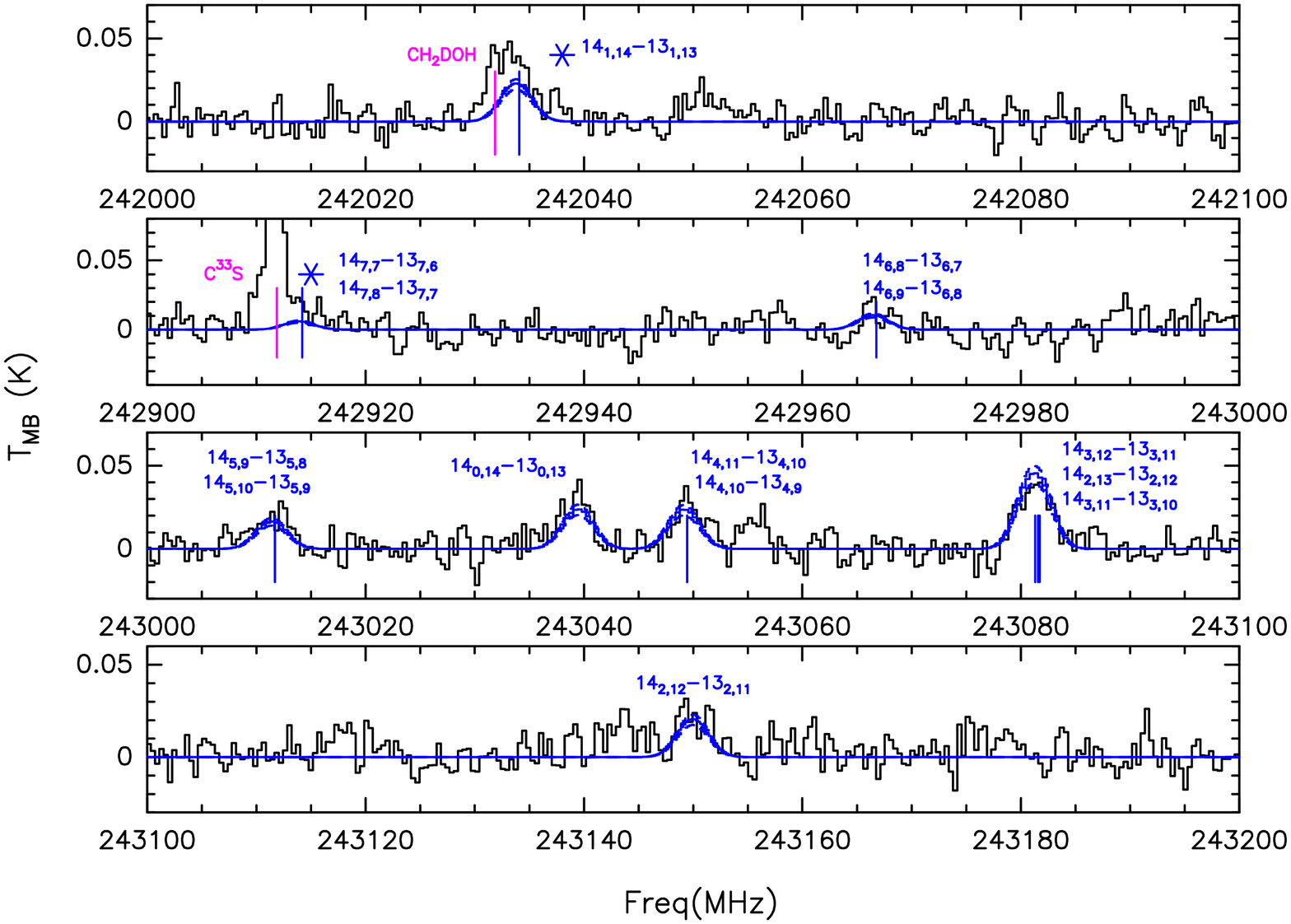}
\includegraphics[angle=0,width=15cm]{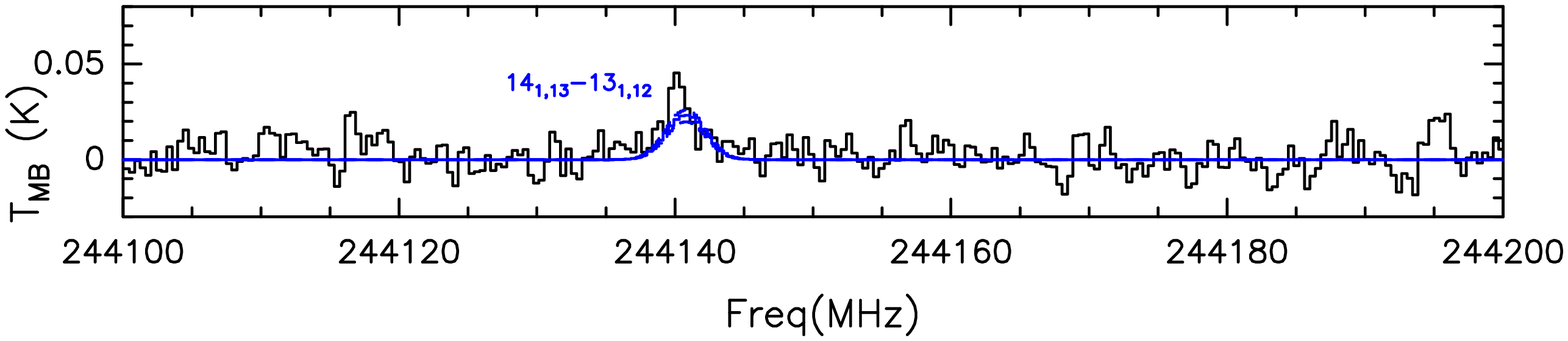}
  \caption{\textit{Continued.}}
  \end{center}
\end{figure*}

\end{document}